\pdfoutput=1

\documentclass[11pt,twoside,a4paper,cmspaper,final,collab]{cms-tdr}
\def\svnVersion{404425}\def\cmsCernNoTag{CERN-EP/2016-205}\def\cmsCernDate{\today}\def\cmsMessage{Published in the European Physical Journal C as \href{http://dx.doi.org/10.1140/epjc/s10052-017-4730-z}{\doi{10.1140/epjc/s10052-017-4730-z.}}}

\begin{document}\cmsNoteHeader{SMP-14-014}

\hyphenation{had-ron-i-za-tion}
\hyphenation{cal-or-i-me-ter}
\hyphenation{de-vices}
\RCS$Revision: 404416 $
\RCS$HeadURL: svn+ssh://svn.cern.ch/reps/tdr2/papers/SMP-14-014/trunk/SMP-14-014.tex $
\RCS$Id: SMP-14-014.tex 404416 2017-05-12 18:03:32Z alverson $
\providecommand{\PV}{\ensuremath{\mathrm{V}}\xspace}
\newlength\cmsFigWidth
\ifthenelse{\boolean{cms@external}}{\setlength\cmsFigWidth{0.85\columnwidth}}{\setlength\cmsFigWidth{0.65\textwidth}}
\ifthenelse{\boolean{cms@external}}{\providecommand{\cmsLeft}{top\xspace}}{\providecommand{\cmsLeft}{left\xspace}}
\ifthenelse{\boolean{cms@external}}{\providecommand{\cmsRight}{bottom\xspace}}{\providecommand{\cmsRight}{right\xspace}}
\ifthenelse{\boolean{cms@external}}{\providecommand{\cmsTable}[1]{#1}}{\providecommand{\cmsTable}[1]{\resizebox{\textwidth}{!}{#1}}}
\providecommand{\exper}{\ensuremath{\,\text{(exp)}}\xspace}
\cmsNoteHeader{SMP-14-014}
\title{Measurement of the WZ production cross section
in pp collisions at $\sqrt{s} = 7$ and 8\TeV and search
for anomalous triple gauge couplings at $\sqrt{s} = 8\TeV$}
\titlerunning{WZ production cross section in pp collisions at 7 and 8\TeV and aTGCs}

\date{\today}

\abstract{
The WZ production cross section is measured by the CMS experiment at
the CERN LHC in proton-proton collision data samples corresponding to integrated luminosities
of 4.9\fbinv collected at $\sqrt{s} = 7\TeV$, and 19.6\fbinv
at $\sqrt{s} = 8\TeV$. The measurements are performed using the fully-leptonic WZ
decay modes with electrons and muons in the final state.
The measured cross sections for $71 < m_{\Z} < 111\GeV$ are
$\sigma({\Pp\Pp}\to{\PW\Z};~\sqrt{s} = 7\TeV) = 20.14 \pm 1.32\stat \pm  0.38\thy \pm 1.06\exper \pm 0.44\lum$\unit{pb}
and
$\sigma({\Pp\Pp}\to{\PW\Z};~\sqrt{s} = 8\TeV) = 24.09 \pm 0.87\stat \pm  0.80\thy \pm 1.40\exper \pm 0.63\lum$\unit{pb}.
 Differential cross sections with respect to the \cPZ\ boson \pt, the leading jet \pt,
 and the number of jets are obtained using the $\sqrt{s} = 8\TeV$ data. The results
 are consistent with standard model predictions and constraints on anomalous triple
 gauge couplings are obtained.
}

\hypersetup{%
pdfauthor={CMS Collaboration},%
pdftitle={Measurement of the WZ production cross section
in pp collisions at sqrt(s) = 7 and 8 TeV and search for anomalous
triple gauge couplings at sqrt(s) = 8 TeV},%
pdfsubject={CMS},%
pdfkeywords={CMS, physics}}

\maketitle

\section{Introduction}

The measurement of the production of electroweak heavy vector boson pairs
(diboson production) in proton-proton collisions represents an important
test of the standard model (SM) description of electroweak
and strong  interactions at the \TeV scale. Diboson production is
sensitive to the self-interactions between electroweak gauge
bosons as predicted by the
$\mathcal{SU}(2)_{L} \times \mathcal{U}(1)_{Y}$
gauge structure of electroweak interactions.
Triple and quartic gauge couplings (TGCs and
QGCs) can be affected by new physics
phenomena involving new particles at higher energy scales.
The $\PW\cPZ$  cross section measured in this paper is sensitive to WWZ
couplings, which are non-zero in
the SM. WZ production also represents an important background in several
searches for physics beyond the SM, such as the search for the
SM Higgs boson~\cite{Chatrchyan:2013iaa}, searches for new
resonances~\cite{Khachatryan:2014xja, Aad:2014pha}, or
supersymmetry~\cite{Khachatryan:2014qwa,Chatrchyan:2014aea,
Aad:2016tuk,Aad:2011cwa}.

\begin{figure*}[hbt]
  \centering
    \includegraphics[width=0.25\textwidth]{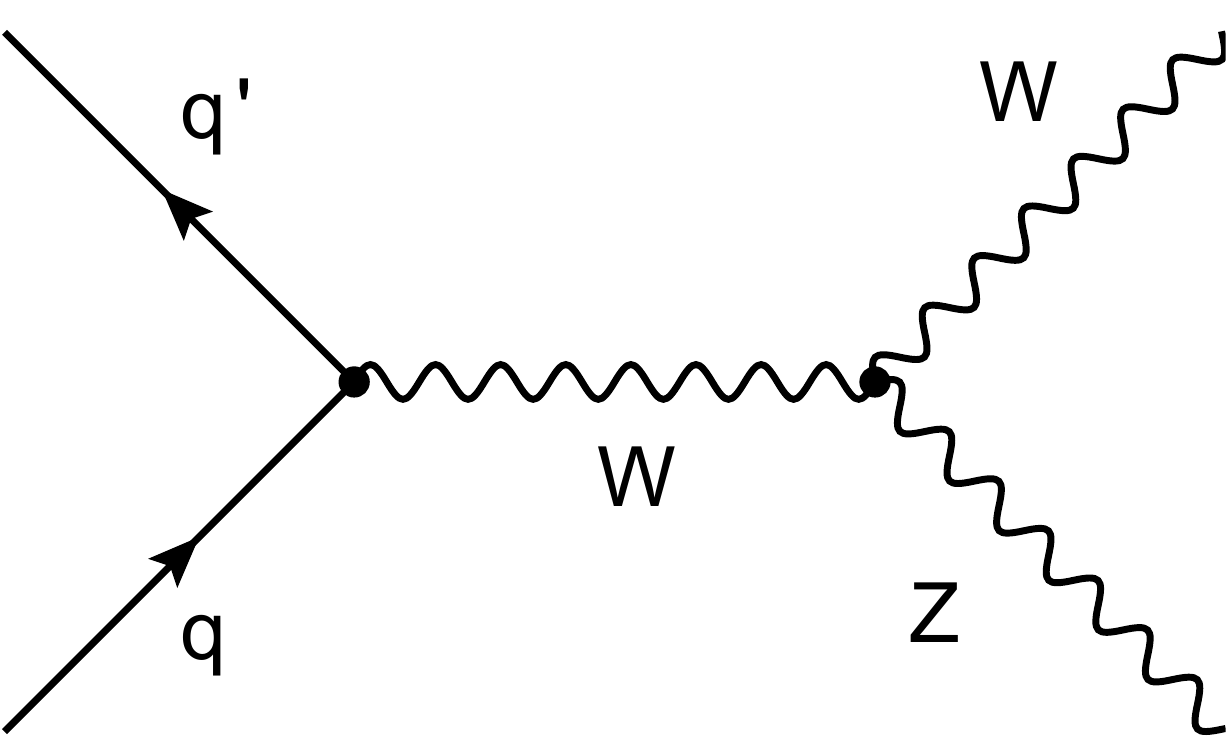}\hfil
    \includegraphics[width=0.25\textwidth]{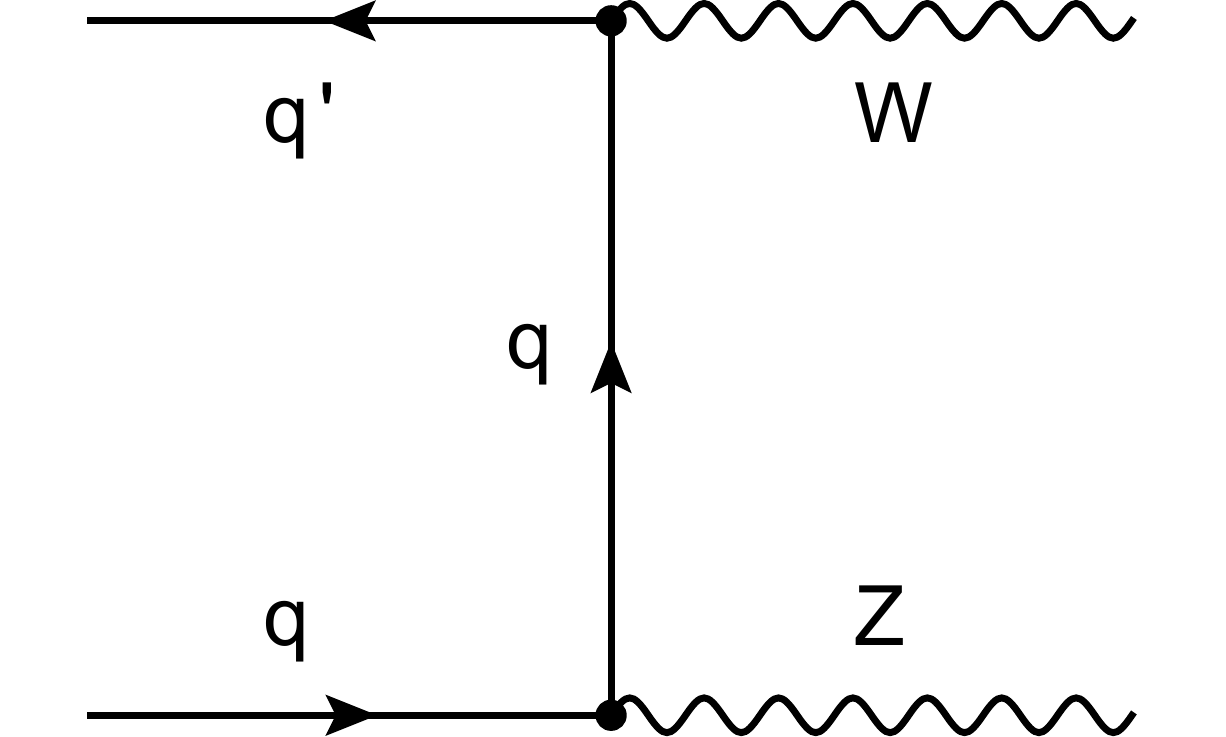}\hfil
    \includegraphics[width=0.25\textwidth]{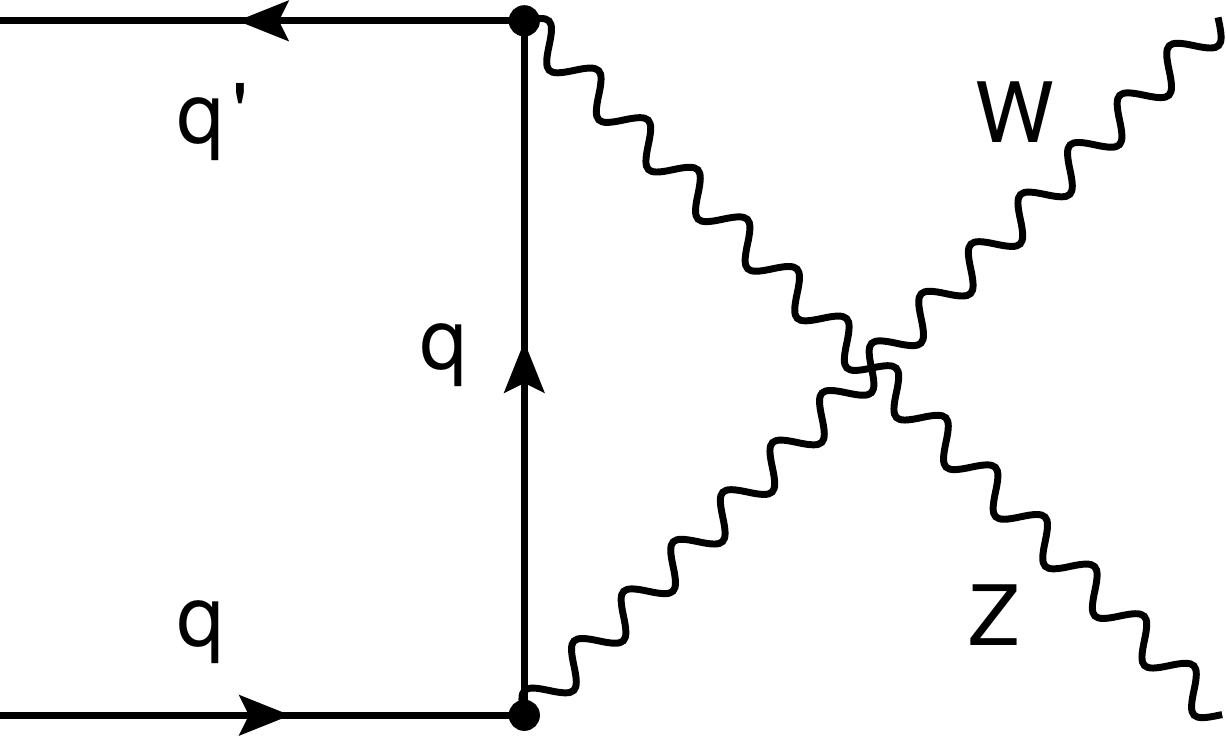}
    \caption{Leading-order Feynman diagrams for WZ production in
      proton-proton collisions. The three diagrams represent contributions from
      (left) $s$-channel through TGC, (middle) $t$-channel, and (right) $u$-channel.
\label{fig:wzFeynman}}
\end{figure*}

We present a study of WZ production in proton-proton collisions
based on data recorded by the CMS detector at the CERN LHC in 2011 and
2012, corresponding to integrated luminosities of 4.9\fbinv collected at
$\sqrt{s}=7\TeV$, and 19.6\fbinv collected at $\sqrt{s}=8\TeV$.
The measurements use purely leptonic
final states in which the Z boson decays into a pair of electrons or muons,
and the W boson decays into a neutrino and an electron or a muon.
At leading order (LO) within the SM, WZ production in proton-proton collisions occurs
through quark-antiquark interactions in the $s$-, $t$-, and $u$-channels, as illustrated
by the Feynman diagrams shown in Fig.~\ref{fig:wzFeynman}.
Among them, only the $s$-channel includes a TGC vertex.
Our measured final states also include contributions from diagrams where the $\cPZ$ boson
is replaced with a virtual photon ($\gamma^*$) and thus include $\PW\gamma^*$ production.
We refer to the final states as $\PW\cPZ$ production because the $\PZ$ contribution is
dominant for the phase space of this measurement.
Hadron collider WZ production
has been previously observed at both the
Tevatron~\cite{Abazov:2012cj,Aaltonen:2012vu} and the
LHC~\cite{Aad:2012twa,Aad:2016ett,
Aaboud:2016yus,Khachatryan:2016tgp,
Aad:2014mda, CMS_WV7TeV:aTGC}.

We first describe measurements of the inclusive WZ production
cross section at both centre-of-mass energies.
The measurements are restricted to the phase space in which the invariant mass
of the two leptons from the Z boson decay lies within 20\GeV of the nominal
Z boson mass~\cite{Agashe:2014kda}.
Using the larger integrated luminosity collected at $\sqrt{s}=8\TeV$, we
also present measurements of the differential cross section as
a function of the Z boson transverse momentum \pt, the number of
jets produced in association with the $\PW\cPZ$  pair, and the \pt
of the leading associated jet. The measurements involving jets are
especially useful for probing the contribution of higher-order
QCD processes to the cross section.

Finally, we present a search for anomalous WWZ couplings based
on a measurement of the \pt spectrum of the Z boson.
The search is formulated both in the framework of anomalous
couplings and in an effective field theory approach.

\section{The CMS detector}

The central feature of the CMS apparatus is a superconducting solenoid of 6\unit{m}
internal diameter, providing a magnetic field of 3.8\unit{T}. Within the solenoid volume are a silicon pixel and strip tracker, a lead tungstate crystal
electromagnetic calorimeter (ECAL), and a brass and scintillator hadron calorimeter
(HCAL), each composed of a barrel and two endcap sections. Muons are measured in
gas-ionization detectors embedded in the steel flux-return yoke outside the solenoid
with detection planes made using three technologies: drift tubes, cathode strip
chambers, and resistive-plate chambers. Extensive forward calorimetry complements
the coverage provided by the barrel and endcap detectors. The silicon tracker measures
charged particles within the pseudorapidity range $\abs{\eta}< 2.50$. The ECAL provides
coverage in $\abs{ \eta }< 1.48 $ in a barrel region and
$1.48 <\abs{\eta} < 3.00$ in two endcap regions. Muons are measured in the range $\abs{\eta}< 2.40$.

A more detailed description of the CMS detector, together with a definition of the
coordinate system used and the relevant kinematic variables, can be found in
Ref.~\cite{Chatrchyan:2008zzk}.

\section{Simulated samples}

Several Monte Carlo (MC) event generators are used to simulate signal and
background processes.
The W($\cPZ/\gamma^*$) signal for $m_{\cPZ/\gamma^*} > 12 \GeV$ is generated at LO with
\MADGRAPH 5.1~\cite{Madgraph5} with up to two additional
partons at matrix element level.
The $\ttbar$, tW, and $\qqbar \to \Z\Z$ processes are
generated at next-to-leading order (NLO) with
\POWHEG 2.0~\cite{Alioli:2008gx,Nason:2004rx,Frixione:2007vw}.
The ${\Pg\Pg \to \Z\Z}$ process is simulated at leading order (one loop)
with \textsc{gg2zz} \cite{Binoth:2008pr}.
Other background processes are generated at
LO with \MADGRAPH and include
\Z+jets, $\PW\gamma^*$ (with $m_{\gamma*}<12\GeV$),
$\Z\gamma$ as well as processes  with at least three bosons in
the decay  chain comprised of $\PW\Z\Z$, $\Z\Z\Z$, $\PW\PW\Z$, $\PW\PW\PW$, $\ttbar\PW$,
$\ttbar\Z$, $\ttbar\PW\PW$, $\ttbar\gamma$ and
$\PW\PW\gamma$, collectively referred to as VVV.
For the modeling of anomalous triple gauge couplings (aTGCs), the NLO \MCFM 6.3~\cite{MCFM}
Monte Carlo program  is used to compute weights that are applied
to the WZ signal sample generated with  \MADGRAPH.  In all samples,
the parton-level events are interfaced with \PYTHIA 6.426~\cite{Sjostrand:2006za} to
describe parton showering,  hadronization, fragmentation, and the
underlying event
 with the Z2* tune~\cite{Field:2010bc}.
For LO generators, the default set of parton
distribution functions (PDFs) used is CTEQ6L1~\cite{CTEQ66},
while NLO CT10~\cite{ct10} is used with NLO generators.
For all processes, the detector
response is simulated with a detailed description of the CMS detector, based on
the \GEANTfour~package~\cite{GEANT}. The event reconstruction is performed with
the same algorithms as are used for data. The simulated samples include additional
interactions per bunch crossing (pileup). Simulated events are weighted so
the pileup distribution in the simulation matches the one observed in data.

\section{Event reconstruction and object identification}

The measurement of the $\PW\Z\to\ell\nu\ell'\ell'$ decay, where
$\ell,\ell'=\Pe$ or $\mu$, relies on the effective identification of electrons and muons, and
an accurate measurement of missing transverse momentum.
The lepton selection requirements used in this measurement are the
same as those used in the Higgs boson $\PH\to{\PW\PW} \to \ell \ell^\prime\nu\nu$
measurement~\cite{Chatrchyan:2013iaa}. The kinematic properties of the
final-state leptons in those two processes are very similar and the two
measurements are affected by similar sources of lepton backgrounds.

Events are required to be accepted by one of the following double-lepton triggers:
two electrons or two muons with transverse momentum thresholds of 17\GeV for the
leading lepton, and 8\GeV for the trailing one.
For the 8\TeV data sample, events are also accepted when an electron-muon pair
satisfies the same momentum criteria.

A particle-flow (PF) algorithm~\cite{CMS-PAS-PFT-09-001,CMS-PAS-PFT-10-001} is used
to reconstruct and identify each individual particle with
an optimized combination of information from the various elements of the CMS detector:
clusters of energy deposits measured by the calorimeters, and charged-particle tracks
identified in the central tracking system and the muon detectors.

Electrons are reconstructed by combining information from
the ECAL and tracker~\cite{
Khachatryan:2015hwa}. Their identification relies on a multivariate regression technique that combines
observables sensitive to the amount of bremsstrahlung along the electron trajectory,
the geometrical and momentum matching between the electron trajectory in the tracker
and the energy deposit in the calorimeter, as well as the shower
shape. Muons are reconstructed using
information from both the tracker and the muon
spectrometer~\cite{CMS-PAS-MUO-10-002}. They
must satisfy requirements on the number of hits in the
layers of the tracker and in the muon spectrometer, and on the
quality of the full track fit. All lepton candidates are required to be
consistent with the primary vertex of the event, which is chosen as
the vertex with the highest $\sum \pt^2$ of its associated tracks.
This criterion provides the correct assignment for the primary
vertex in more than 99\% of both signal and background events for the pileup
distribution observed in data. Both electrons and muons are required
to have $\pt>10\GeV$. Electrons (muons) must satisfy $\abs{\eta}<2.5$ (2.4).

Charged leptons from $\PW$ and $\cPZ$ boson decays are mostly isolated from other
final-state particles in the event. Consequently, the selected leptons are required
to be isolated from other activity in the event to reduce the backgrounds from
hadrons that are misidentified as leptons or from leptons produced in hadron decays
when they occur inside or near hadronic jets.
The separation between two reconstructed objects in
the detector is measured with the variable
$\Delta R = \sqrt{ (\Delta \eta)^2 + (\Delta \phi)^2}$, where $\phi$ is the azimuthal angle.
To measure the lepton isolation, we consider a $\Delta R=0.3$ cone
around the lepton candidate track direction at the event vertex.
An isolation variable is then built as the scalar \pt sum of all
PF objects consistent with the chosen primary vertex,
and contained within the
cone. The contribution from the lepton candidate itself is excluded.
For both electrons and muons a correction is applied to account for
the energy contribution in the isolation cone due to pileup.
In the case of electrons, the average energy density in the isolation cone
due to pileup is determined event-by-event and is used to
correct the isolation variable~\cite{Cacciari:subtraction}.
For muons, the pileup contribution from neutral particles to the
isolation is estimated using charged particles associated with pileup
interactions. This isolation variable is required to be smaller
than about 10\% of the candidate lepton \pt. The exact
threshold value depends on the lepton flavour and detector region,
and also on the data taking period: for 7\TeV data, it is 13\%\,(9\%) for
electrons measured in the ECAL barrel (endcaps) and 12\% for muons,
while for 8\TeV data it is 15\% for all electrons. For muons, a
modified strategy
has been used for 8\TeV data to account for the higher pileup
conditions in order to reduce the dependence of this variable
on the number of pileup interactions. It uses a multivariate
algorithm based on the \pt sums of particles
around the lepton candidates built for $\Delta R$ cones of different
sizes~\cite{Chatrchyan:2013iaa}.

The lepton reconstruction and selection efficiencies
and associated uncertainties are determined using a tag-and-probe
method with $\Z\to\ell\ell$ events~\cite{wzxs} chosen using
the same criteria in data and simulation in several
(\pt,$\eta$) bins.  Ratios of efficiencies from data and simulation
are calculated for each bin.
To account for differences between data and
simulation, the simulated samples are reweighted by these ratios
for each selected lepton in the event.
The total uncertainty for the lepton efficiencies, including effects
from trigger, reconstruction, and selection
amounts to roughly 2\% per lepton.
The lepton selection criteria in the 7~and 8\TeV
samples are chosen to maintain a stable efficiency
throughout each data sample.

Jets
are reconstructed  from PF objects
using the anti-\kt  clustering
algorithm~\cite{Cacciari:2008gp, Cacciari:2011ma}
with a size parameter $R$ of 0.5.
The energy of photons is
obtained from the ECAL measurement. The energy of electrons is determined from a
combination of the electron momentum at the primary interaction vertex as determined
by the tracker, the energy of the corresponding ECAL cluster, and the energy sum of
all bremsstrahlung photons spatially compatible with origination from the electron
track. The energy of muons is obtained from the curvature of the corresponding track.
The energy of charged hadrons is determined from a combination of their momentum
measured in the tracker and the matching ECAL and HCAL energy deposits, corrected for
the response function of the calorimeters to hadronic showers. Finally, the energy of
neutral hadrons is obtained from the corresponding corrected ECAL and HCAL energy.
The jet momentum is determined as the vector sum of all particle
momenta in the jet.
A correction is applied to jet energies to take into account
the contribution from pileup.
Jet energy corrections are derived from the simulation, and are
confirmed with in situ measurements with the energy balance of
dijet and photon + jet events~\cite{Chatrchyan:2011ds}. The jet
energy resolution amounts typically to 15\% at 10\GeV, 8\% at 100\GeV,
and 4\% at 1\TeV.
Additional  selection criteria are applied to each event to
remove  spurious jet-like features originating from isolated noise
patterns  in certain HCAL regions.

The missing transverse momentum vector \ptvecmiss is defined as
the negative vector sum of the transverse momenta of all reconstructed particles
in an event. Its magnitude is referred to as \ETmiss.

\section{Event selection and background estimates}

We select $\PW\cPZ\to\ell\nu\ell'\ell'$ decays with
$\PW \to\ell\nu$ and $\cPZ\to\ell'\ell'$, where $\ell$ and $\ell'$ are
electrons or muons. These decays are characterized by a pair of
same-flavour, opposite-charge, isolated leptons with an invariant mass consistent
with a Z boson, together with a third isolated lepton and a
significant amount of missing transverse energy \MET associated with the escaping
neutrino. We consider four different signatures corresponding to the
flavour of the leptons in the final state: $\Pe\Pe\Pe$, $\Pe\Pe\mu$, $\Pe\mu\mu$ and
$\mu\mu\mu$.

The four final states are treated independently for the cross section
measurements and for the search for anomalous couplings, and are
combined only at the level of the final results.
Unless explicitly stated otherwise, identical selection criteria are applied to the 7 and 8\TeV samples.

Candidate events are triggered by requiring the presence of two
 electrons or two muons. In the 8\TeV sample, events
triggered by the presence of an electron and a muon are also accepted.
The trigger efficiency for signal-like events that pass the event selection is
measured to be larger than 99\%.
The candidate events are required to contain exactly three leptons matching all
selection criteria.  In the 8\TeV analysis, the invariant mass of the three leptons
is required to be larger than 100\GeV.
The \cPZ\ boson candidates are built from two oppositely charged, same-flavour, isolated leptons.
The leading lepton is required to have $\pt > 20\GeV$.
The Z boson candidate invariant mass should lie within
20\GeV of the nominal Z boson mass: $71 < m_{\ell\ell} < 111\GeV$. If two matching pairs are found, the Z
boson candidate with the mass closest to the nominal \cPZ\ boson mass is selected. The remaining
lepton is associated with the W boson and is required to have
 $\pt > 20\GeV$ and to be separated from both leptons in the \cPZ\ boson decay
by $\Delta R > 0.1$. Finally, to account for the escaping
neutrino, \MET is required to be larger than 30\GeV.

Background sources with three reconstructed leptons include events with prompt
leptons produced at the primary vertex or leptons from displaced vertices, as
well as jets.

The background contribution from nonprompt leptons, dominated by
$\ttbar$ and Z+jets events in which one of the three
reconstructed leptons is misidentified, is estimated using a
procedure similar to Ref.~\cite{CMSWWxsTGC35pb}.
In this procedure, the amount of background in the signal region is
estimated using the yields observed in several mutually exclusive samples
containing events
that did not satisfy some of the lepton selection requirements.
The method uses the distinction between a loose and a tight lepton
selection. The tight selection is identical to the one used in the
final selection, while some of the lepton
identification requirements used in the final selection are
relaxed in the loose selection. The procedure starts from a sample,
called the loose sample, with three leptons passing loose identification criteria and
otherwise satisfying all other requirements of the WZ selection.
This sample receives contributions from events with three prompt (p)
leptons, two prompt leptons and one nonprompt (n) lepton, one prompt lepton
and two nonprompt leptons, and three nonprompt leptons. The event yield
of the loose sample $N_\mathrm{LLL}$ can thus be expressed as,
\begin{equation}
  N_\mathrm{LLL} = n_\mathrm{ppp} + n_\mathrm{ppn} + n_\mathrm{pnp} + n_\mathrm{npp}
+ n_\mathrm{nnp} + n_\mathrm{npn} + n_\mathrm{pnn} + n_\mathrm{nnn}.
\label{eq:looseSample}
\end{equation}
In this expression, the first, second and third indices refer to
the leading and subleading leptons from the Z boson decay and to the lepton
from the W boson decay, respectively.
The loose sample can be divided into subsamples depending on whether
each of the three leptons
passes or fails the tight selection. The number of events in each subsample
is labeled $N_{ijk}$ with $i,j,k = \mathrm{T,F}$ where T and F stand for
leptons passing or failing the tight selection, respectively.
The yield in each of these subsamples can be
expressed as a linear combination of the unknown yields
$n_{\alpha\beta\gamma}$ ($\alpha,\beta,\gamma \, \in \{ \mathrm{p,n} \}$),
\begin{equation}
N_{ijk} = \sum_{\alpha,\beta,\gamma \, \in \{ \mathrm{p,n} \}}
C^{ijk}_{\alpha\beta\gamma}  n_{\alpha\beta\gamma}, \quad
 i,j,k=\mathrm{T,F},
\label{eq:matrix}
\end{equation}
where the coefficients $C^{ijk}_{\alpha\beta\gamma}$ depend
on  the efficiencies $\epsilon_\mathrm{p}$ and
$\epsilon_\mathrm{n}$, which stand for the probabilities
of prompt and nonprompt leptons, respectively, to pass the tight lepton selection
provided they have passed the loose selection.
For example, starting from Eq.~(\ref{eq:looseSample}), the number of
events with all three leptons passing the tight selection $N_\mathrm{TTT}$ can be written
as
\begin{multline}
N_\mathrm{TTT} =
                n_\mathrm{ppp}\epsilon_\mathrm{p_1}\epsilon_\mathrm{p_2}\epsilon_\mathrm{p_3} +
                n_\mathrm{ppn}\epsilon_\mathrm{p_1}\epsilon_\mathrm{p_2}\epsilon_\mathrm{n_3} +
                n_\mathrm{pnp}\epsilon_\mathrm{p_1}\epsilon_\mathrm{n_2}\epsilon_\mathrm{p_3} \\
               + n_\mathrm{npp}\epsilon_\mathrm{n_1}\epsilon_\mathrm{p_2}\epsilon_\mathrm{p_3} +
                n_\mathrm{nnp}\epsilon_\mathrm{n_1}\epsilon_\mathrm{n_2}\epsilon_\mathrm{p_3} +
                n_\mathrm{npn}\epsilon_\mathrm{n_1}\epsilon_\mathrm{p_2}\epsilon_\mathrm{n_3} \\
               + n_\mathrm{pnn}\epsilon_\mathrm{p_1}\epsilon_\mathrm{n_2}\epsilon_\mathrm{n_3} +
                n_\mathrm{nnn}\epsilon_\mathrm{n_1}\epsilon_\mathrm{n_2}\epsilon_\mathrm{n_3}.
\end{multline}
The goal is to determine the number of events with three prompt
leptons in the TTT sample, corresponding exactly to the
selection used to perform the measurement.
This yield is
$n_\mathrm{ppp}\epsilon_\mathrm{p_1} \epsilon_\mathrm{p_2}  \epsilon_\mathrm{p_3} $.
The number of events with three prompt leptons in the loose sample,
$n_\mathrm{ppp}$, is obtained by solving the set of linear
equations~(\ref{eq:matrix}).

Independent samples are used to measure the efficiencies
 $\epsilon_\mathrm{p}$ and $\epsilon_\mathrm{n}$~\cite{CMSWWxsTGC35pb}.
The prompt lepton efficiency $\epsilon_\mathrm{p}$ is obtained from
a $\Z\to \ell\ell$ sample,
while the nonprompt lepton efficiency $\epsilon_\mathrm{n}$ is
measured using a quantum chromodynamics (QCD) multijet sample.
Events in this sample are triggered by a single lepton.
The lepton selection used in these triggers
is looser than the loose lepton selection referred to earlier
in this section. The leading jet in the event is required to
be well separated from the triggering lepton and have
a transverse momentum larger than 50\GeV for the 7\TeV
data sample, and larger than 35\,(20)\GeV for the 8\TeV sample
if the triggering lepton is an electron (muon).
Events with leptons from \Z decays are rejected by requiring exactly
one lepton in the final state. To reject events with leptons from \PW\ decays,
both the missing transverse energy and the \PW\ transverse mass are required
to be less than 20\GeV.
This selection provides a clean sample to estimate the nonprompt
lepton efficiency. Both efficiencies  $\epsilon_\mathrm{p}$ and
$\epsilon_\mathrm{n}$ are measured in several lepton $(\pt, \eta)$ bins.
For 7\TeV\,(8\TeV) data, the measured nonprompt efficiencies
for leptons are in the range 1--6\%\,(1--10\%), while they are
in the range 1--5\%\,(7--20\%) for muons. The measured
prompt efficiencies lie between 60 and 95\% for electrons, and
between 71 and 99\% for muons for both
the 7 and 8\TeV data samples.

The number of events with
nonprompt leptons in each final state obtained with this method
is given in Table~\ref{tab:yields}. While these results include
the contribution of events with any number of misidentified leptons,
simulation studies show that the contribution from backgrounds
with two or three misidentified leptons, such as W+jets or QCD multijet processes,
is negligible, so the nonprompt lepton background is completely dominated
by $\ttbar$ and Z+jets processes.

The remaining background is composed of events with three
prompt leptons, such as the \mbox{$ZZ\to 2\ell 2\ell'$} process in which
one of the four final-state leptons has not been identified, as
well as processes with three or more heavy bosons in the
final states ($VVV$), and the $\PW\gamma^*$ process, with $\gamma^*\to\ell^+\ell^-$.
These backgrounds are estimated from simulation.
The relevant W$\gamma^*$ process  is defined for
low $\gamma^*$ masses, $m_{\gamma^*} < 12\GeV$, so
it does  not overlap with the W$\gamma^*$ process included in the signal
simulation and it is simulated separately. It is considered a background
since it does not fall in the fiducial phase space of the proposed measurement.
Such W$\gamma^*$ processes would be accepted by the  event
selection only if the charged lepton from the W decay is wrongly
interpreted as coming from the $\Z/\gamma^*$ decay.
The contribution of $\Z\gamma$ events in which the photon
is misidentified as a lepton is also determined from simulation.
Prompt photons will not contribute to a nonprompt lepton signal
since photons and electrons have a similar signature in the detector.
Prompt photons in $\Z\gamma$ events will also typically be
isolated from other final state particles.

We finally consider the contribution of WZ decays, in which either the
$\PW$ or $\cPZ$ boson decays to a $\tau$ lepton. Such decays are considered
a background to the signal. Their contribution is subtracted
using the fraction of selected WZ decays that have  $\tau$ leptons
in the final state. This fraction, labeled $f_\tau$,  is estimated from simulation for
each of the four final states, and lies in between 6.5 and 7.6\%. This
background is almost entirely composed of WZ events with $\PW\to\tau\nu$
decays where the $\tau$ lepton subsequently decays into an electron or a muon.

\begin{figure}[hbtp]
  \centering
    \includegraphics[width=0.49\textwidth]{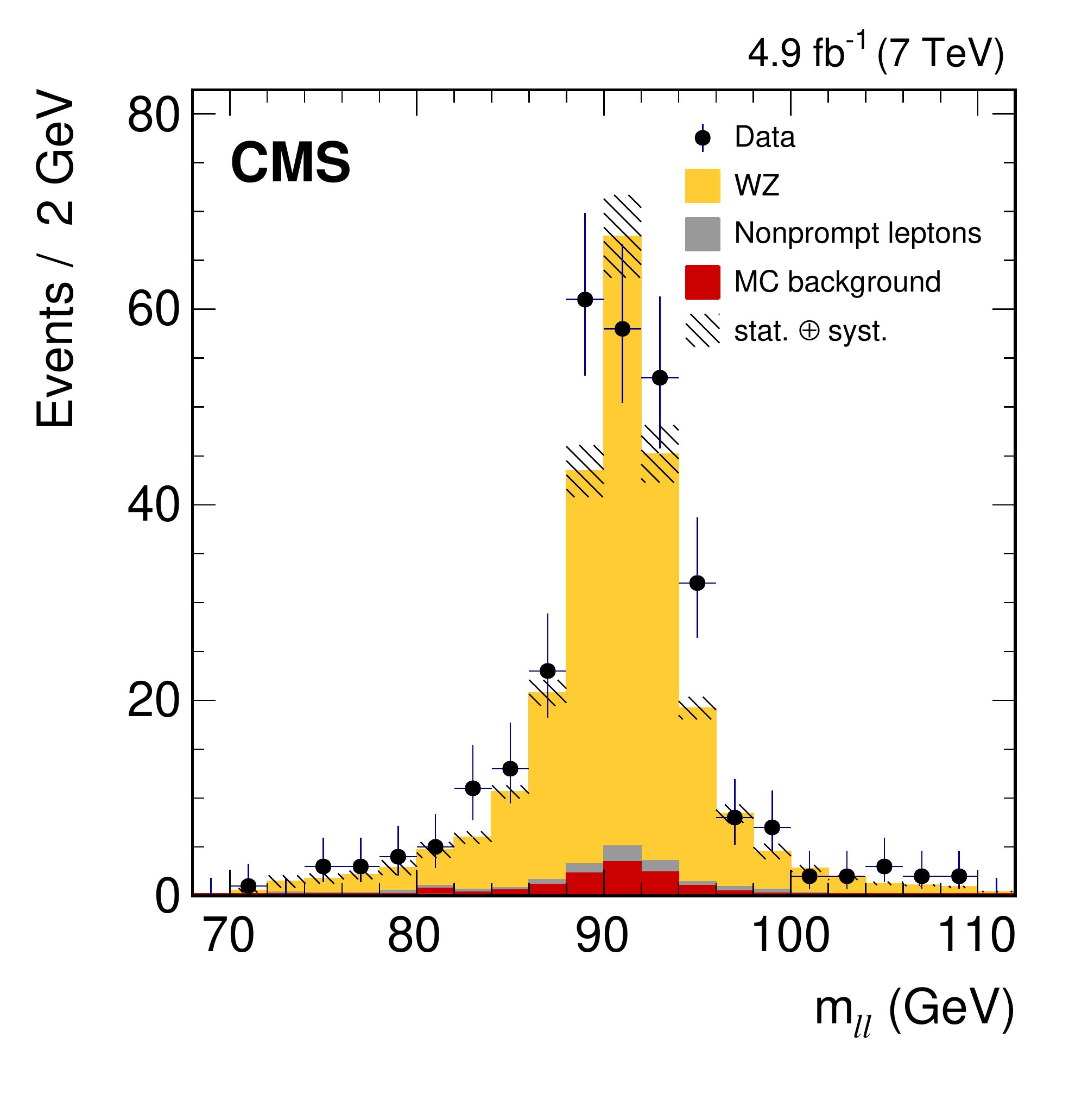}
    \includegraphics[width=0.49\textwidth]{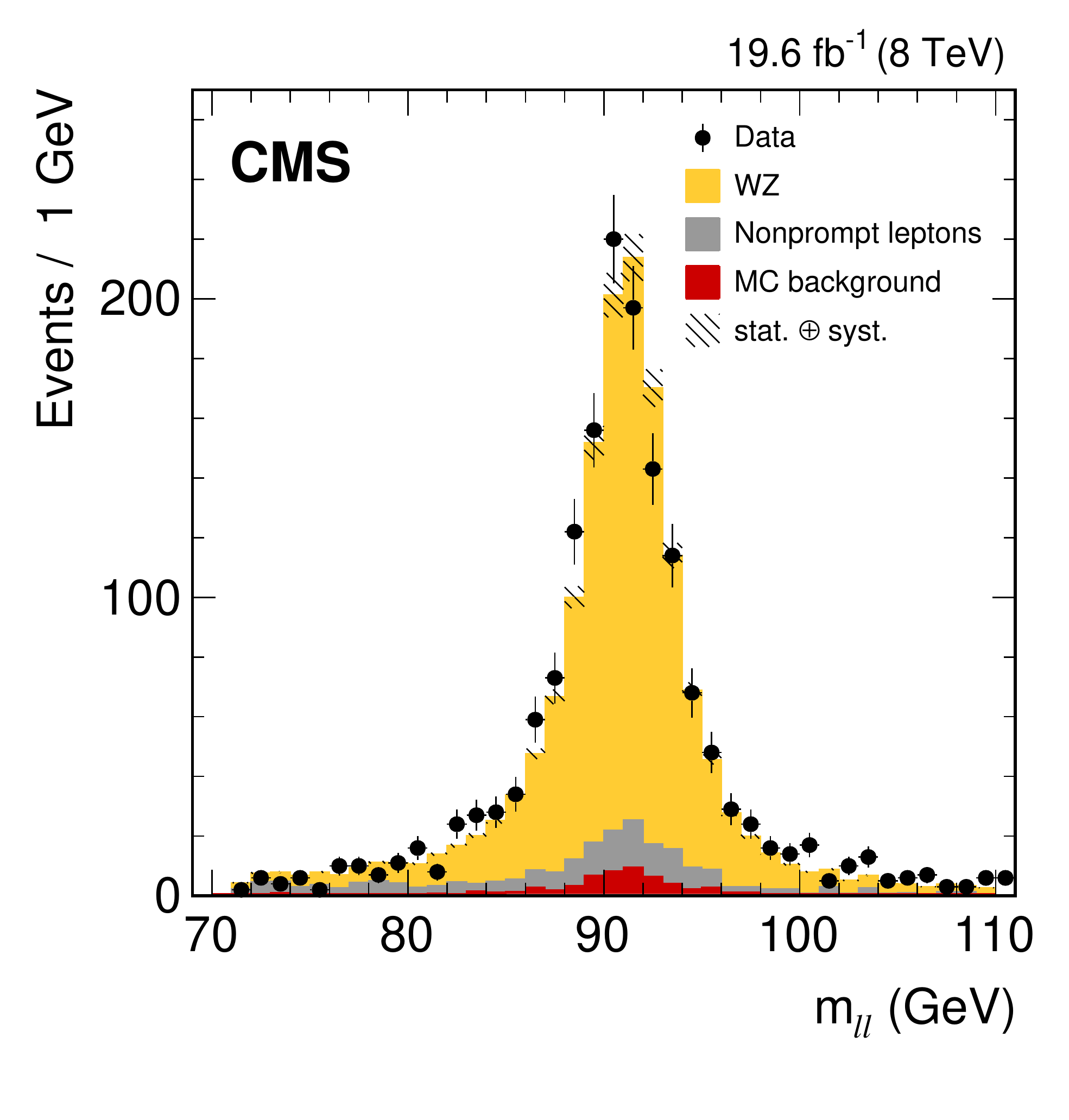}
    \caption{    Distributions of the dilepton invariant mass $m_{\ell\ell}$ in the WZ
    candidate events in 7\TeV  (\cmsLeft) and 8\TeV (\cmsRight) data.
    Points represent  data and the shaded histograms represent the WZ
    signal and the  background processes.
    The contribution from nonprompt leptons, dominated by the \ttbar   and Z+jets production, is obtained from data control
    samples.  The contribution from all other backgrounds,
    labeled `MC background', as well as the signal contribution are
    determined from simulation. }
    \label{fig:invMass2Lep}
\end{figure}

After applying all selection criteria, 293\,(1559) events are selected from the
7\,(8)\TeV data corresponding to an integrated luminosity of 4.9\,(19.6)\fbinv. The yields for each
leptonic channel, together with the expectations from MC simulation and
data control samples
are given in Table~\ref{tab:yields}. The inclusive
distributions of the dilepton invariant mass $m_{\ell\ell}$ for both 7 and
8\TeV data samples are shown in Fig.~\ref{fig:invMass2Lep}.
\begin{table*}[htb]
\centering
\topcaption{Expected and observed event yields at $\sqrt{s} = 7$ and 8\TeV.
The contributions from \ttbar, Z+jets, and other processes with
nonprompt leptons have been determined from data control samples,
as described in the text. Backgrounds with at least three bosons in
the decay  chain comprised of WZZ, $\Z\Z\Z$, WWZ, WWW, $\ttbar\PW$,
$\ttbar \Z$, $\ttbar\PW\PW$, $\ttbar\gamma$ and
$\PW\PW\gamma$ events, are referred to as $\PV\PV\PV$.
Combined statistical and systematic uncertainties are shown, except for
the WZ signal where only statistical uncertainties are shown.
\label{tab:yields}}
\cmsTable{
\begin{tabular}{lccccc}
\hline
Sample              &             $\Pe\Pe\Pe$  &                $\Pe\Pe\mu$  &              $\mu\mu\Pe$  &            $\mu\mu\mu$  &                    Total  \\ \hline\\[1ex]
\multicolumn{6}{c}{$\sqrt{s} = 7\TeV$; ${\mathcal{L}} = 4.9\fbinv$} \\
 \hline
Nonprompt leptons  &   2.2 $\pm$ 2.1  &   1.5 $^{+4.8}_{-1.5}$    &   2.4 $^{+5.1}_{-2.4}$   &   1.8 $^{+7.5}_{-1.8}$  &    7.9 $^{+13.0}_{-5.0}$  \\
ZZ                  &   2.0 $\pm$ 0.3  &   3.5 $\pm$ 0.5         &   2.7 $\pm$ 0.4         &   5.1 $\pm$ 0.7         &   13.3 $\pm$ 1.9          \\
Z$\gamma$           &   0              &   0                     &   0.5 $\pm$ 0.5         &   0                     &    0.5 $\pm$ 0.5          \\
VVV                 &   1.6 $\pm$ 0.8  &   2.0 $\pm$ 1.0         &   2.4 $\pm$ 1.2         &   3.0 $\pm$ 1.5         &    9.0 $\pm$ 4.5          \\ \hline
Total background
($N_{\text{bkg}}$)  & 3.8 $\pm$ 2.3    &  6.0 $\pm$ $^{+4.9}_{-1.9}$ & 8.0 $^{+5.1}_{-2.4}$   &   9.9 $^{+7.7}_{-2.4}$    &   30.7 $^{+13.9}_{-7.0}$    \\
WZ                  &  44.7 $\pm$ 0.5  &  49.8 $\pm$ 0.5         &  56.0 $\pm$ 0.5         &  73.8 $\pm$ 0.6         &  224.3 $\pm$ 1.1          \\ \hline
Total expected      &  50.5 $\pm$ 2.3  &  56.8 $^{+5.0}_{-1.9}$  &  64.0 $^{+5.3}_{-2.8}$  &  83.7 $^{+7.7}_{-2.5}$  &  255 $^{+14.0}_{-7.0}$    \\
Data ($N_{\text{obs}}$)  &  64         &  62                     &  70                     &  97                     &  293                      \\ \hline\\[1ex]
\multicolumn{6}{c}{$\sqrt{s} = 8\TeV$; ${\mathcal{L}} = 19.6\fbinv$} \\
\hline
Nonprompt leptons  &   18.4 $\pm$ 12.7  &   32.0 $\pm$ 21.0  &   54.4 $\pm$ 33.0  &   62.4 $\pm$ 37.7  &   167.1 $\pm$ 55.8  \\
ZZ                  &    2.1 $\pm$  0.3  &    2.4 $\pm$  0.4  &    3.2 $\pm$  0.5  &    4.7 $\pm$  0.7  &    12.3 $\pm$  1.0  \\
$\Z\gamma$           &    3.4 $\pm$  1.3  &    0.4 $\pm$  0.4  &    5.2 $\pm$  1.8  &    0               &     9.1 $\pm$  2.2  \\
$\PW\gamma^*$         &    0               &    0               &    0               &    2.8 $\pm$  1.0  &     2.8 $\pm$  1.0  \\
$VVV$                 &    6.7 $\pm$  2.2  &    8.7 $\pm$  2.8  &   11.6 $\pm$  3.8  &   14.8 $\pm$  5.1  &    41.9 $\pm$  7.3  \\ \hline
Total background
($N_{\text{bkg}}$)  &   30.6 $\pm$ 13.0  &   43.5 $\pm$ 21.2  &   74.4 $\pm$  33.3 &   84.7 $\pm$ 38.1  &    233.2 $\pm$ 56.3 \\
WZ                  &  211.1 $\pm$  1.6  &  262.1 $\pm$  1.8  &  346.7 $\pm$  2.1  &  447.8 $\pm$  2.4  &  1267.7 $\pm$  4.0  \\ \hline
Total expected      &  241.6 $\pm$ 13.1  &  305.7 $\pm$ 21.3  &  421.0 $\pm$ 33.3  &  532.4 $\pm$ 38.2  &  1500.8 $\pm$ 56.5  \\
Data ($N_{\text{obs}}$) &  258  &               298               &  435               &  568               &  1559               \\ \hline
\end{tabular}
}
\end{table*}

\section{Systematic uncertainties}

\label{sec:systematics}

Systematic uncertainties can be grouped into three categories:
the determination of signal efficiency, the estimation of
background yields, and the luminosity measurement.

The first group includes uncertainties affecting the signal efficiency,
referred to as $\epsilon_{\rm{sig}}$, which
accounts for both detector geometrical acceptance and reconstruction
and selection efficiencies. It is determined from simulation.
Uncertainties on $\epsilon_{\rm{sig}}$ depend on theoretical uncertainties in the PDFs.
The PDF uncertainty is evaluated following the prescription in Ref.~\cite{pdfunc}
using the CTEQ66~\cite{CTEQ66} PDF set.
The uncertainties from normalization ($\mu_R$) and factorization
($\mu_F$) scales are estimated by varying both scales independently
in the range ($0.5 \mu_0, 2\mu_0$) around their nominal value
$\mu_0=0.5 (M_{\Z}+M_{\PW})$ with the constraint $0.5 \leq \mu_R / \mu_F \leq 2$.
The signal efficiency $\epsilon_{\text{sig}}$ is also affected by experimental
uncertainties in the muon momentum scale and in the
electron energy scale, lepton reconstruction and identification efficiencies,
\MET calibration scale, and pileup contributions. The effect of the muon momentum
scale is estimated by varying the momentum of each muon in the simulated
signal sample within the momentum scale uncertainty, which is 0.2\%~\cite{CMS-PAS-MUO-10-002}.
The same is done for electrons by varying the energy of reconstructed electrons
within the uncertainty of the energy scale measurement, which is \pt and $\eta$
dependent and is typically below 1\%. The signal efficiency $\epsilon_{\text{sig}}$ also
depends on the uncertainties in the ratios of observed-to-simulated
efficiencies of the lepton trigger, reconstruction, and identification
requirements.
These ratios are used in the determination of $\epsilon_{\text{sig}}$
to account for efficiency differences between data and simulation.
They are varied within their uncertainties, which depend
on the lepton \pt and $\eta$ and are about 1\%.
The uncertainty from the \MET calibration is
determined by scaling up and down the energy of all objects
used for the \MET determination within their uncertainties.
Finally, $\epsilon_{\text{sig}}$ is  affected by the uncertainty
in the pileup contribution.
Simulated events are reweighted to match the distribution of
pileup interactions, which is estimated using a procedure that
extracts the pileup from the instantaneous bunch
luminosity and the total inelastic pp cross section.
The weights applied to simulated
events are changed by varying this cross section by 5\%
uncertainty~\cite{Chatrchyan:2012nj}.

The second group comprises uncertainties in the background yield. The
uncertainty in the background from nonprompt leptons~\cite{CMSWWxsTGC35pb}
is estimated by varying the leading jet \pt threshold used to
select the control sample of misidentified leptons, since
the energy of the leading jet determines the composition of the sample.
The uncertainties from other background processes, whose
contributions are determined from simulation, are calculated by
varying their predicted cross sections within uncertainties.
The cross sections are varied by 15\%\,(14\%) for  \cPZ\cPZ,
by 15\%\,(7\%) for $\Z\gamma$, by 50\%\,(50\%) for the $\PV\PV\PV$ processes,
and by 20\% for $\PW\gamma^*$ for the 8\TeV\,(7\TeV) measurements,
based on the uncertainties of the measurements of these
processes~\cite{Chatrchyan:2012sga,Chatrchyan:2013oev,Chatrchyan:2013fya,
Khachatryan:2015kea,Khachatryan:2014ewa}.

Finally, the uncertainty in the measurement of the integrated luminosity is
2.2\,(2.6)\% for 7\,(8)\TeV data~\cite{CMS-PAS-SMP-12-008,CMS-PAS-LUM-13-001}.

A summary of all uncertainties is given in
Table~\ref{tab:systematics-7-8tev}.

\begin{table*}[htb]
\centering
\topcaption{Summary of relative uncertainties,
in units of percent, in the WZ cross section measurement
at 7 and 8\TeV.\label{tab:systematics-7-8tev}}
\begin{tabular}{lcccc{c}@{\hspace*{5pt}}cccc}
\hline
\multirow{2}{*}{Source} & \multicolumn{4}{c}{$\sqrt{s} = 7\TeV$} && \multicolumn{4}{c}{$\sqrt{s} = 8\TeV$} \\ \cline{2-5}\cline{7-10}
                        & $\Pe\Pe\Pe$ & $\Pe\Pe\mu$ & $\mu\mu\Pe$ & $\mu\mu\mu$ && $\Pe\Pe\Pe$ &$\Pe\Pe\mu$ & $\mu\mu\Pe$ & $\mu\mu\mu$ \\ \hline
Renorm. and fact. scales         &  1.3 &  1.3 &  1.3 &  1.3   &&   3.0 & 3.0 &  3.0 & 3.0 \\
PDFs                             &  1.4 &  1.4 &  1.4 &  1.4   &&   1.4 & 1.4 &  1.4 & 1.4 \\
Pileup                           &  0.3 &  0.5 &  1.0 &  0.6   &&   0.2 & 0.4 &  0.3 & 0.2 \\
Lepton and trigger efficiency    &  2.9 &  2.7 &  2.0 &  1.4   &&   3.4 & 2.5 &  2.5 & 3.2 \\
Muon momentum scale              &  --- &  0.6 &  0.4 &  1.1   &&   --- & 0.5 &  0.8 & 1.3 \\
Electron energy scale            &  1.9 &  0.8 &  1.2 &  ---   &&   1.4 & 0.8 &  0.8 & ---\\
\MET                             &  3.7 &  3.4 &  4.3 &  3.7   &&   1.5 & 1.5 &  1.6 & 1.2 \\
ZZ cross section                 &  0.5 &  0.9 &  0.6 &  0.9   &&   0.1 & 0.1 &  0.1 & 0.1 \\
$\Z\gamma$ cross section          &  0.0 &  0.0 &  0.1 &  0.0   &&   0.2 & 0.0 &  0.2 & 0.0 \\
\ttbar and Z+jets                &  2.7 &  6.5 &  6.3 &  6.0   &&   4.6 & 7.2 &  6.1 & 7.7 \\
Other simulated backgrounds      &  0.2 &  0.2 &  0.9 &  0.2   &&   1.0 & 1.1 &  1.1 & 1.0 \\ \hline
Total systematic uncertainty     &  6.1 &  7.8 &  8.1 &  7.2   &&   7.0 & 8.6 &  7.7 & 9.2 \\
Statistical uncertainty          & 13.5 & 13.9 & 13.1 & 11.0   &&   7.7 & 7.2 &  6.4 & 5.2 \\
Integrated luminosity uncertainty &  2.2 &  2.2 &  2.2 &  2.2   &&   2.6 & 2.6 &  2.6 & 2.6 \\ \hline
\end{tabular}
\end{table*}

\section{Results}

\subsection{Inclusive cross section measurement}

The inclusive WZ cross section $\sigma ( \Pp\Pp \to \PW\Z +X)$ in the
$\ell\nu\ell'\ell'$ final state is related to the number of observed events
in that final state, $N_{\text{obs}}$, through the following  expression,
\ifthenelse{\boolean{cms@external}}{
\begin{multline*}
\sigma (\Pp\Pp \to {\PW\Z+X}) \, \mathcal{B} ({\PW}\to \ell\nu) \,
 \mathcal{B} ({\Z}\to \ell'\ell')\\
= \left(1 - f_{\tau}\right)\frac{N_\text{obs} - N_\text{bkg}}{\epsilon_{\text{sig}}\,\mathcal{L}},
\end{multline*}
}{
\begin{equation*}
\sigma (\Pp\Pp \to {\PW\Z+X}) \, \mathcal{B} ({\PW}\to \ell\nu) \,
 \mathcal{B} ({\Z}\to \ell'\ell')
= \left(1 - f_{\tau}\right)\frac{N_\text{obs} - N_\text{bkg}}{\epsilon_{\text{sig}}\,\mathcal{L}},
\end{equation*}
}
where $\mathcal{B} ({\PW}\to \ell\nu)$ and $ \mathcal{B} ({\Z}\to
\ell'\ell')$ are the $\PW$ and $\cPZ$ boson leptonic branching fractions per
lepton species, and
$f_{\tau}$ accounts for the expected fraction of selected $\PW\Z\to
\ell\nu\ell'\ell'$ decays produced through at least one prompt $\tau$ decay
in the final state after removing all other backgrounds.
The number of expected background events is $N_\text{bkg}$, and
the number of signal events is determined by subtracting $N_\text{bkg}$ from the observed
data $N_\text{obs}$. The signal efficiency $\epsilon_{\text{sig}}$
accounts for both detector geometrical acceptance and reconstruction and
selection efficiencies.
It is obtained for each of the four final states using the simulated WZ sample
by calculating the ratio of the number of events passing the full selection
to the number of generated $\PW\Z\to \ell\nu\ell'\ell'$ events with
$71 < m_{\ell'\ell'} < 111\GeV$, where $m_{\ell'\ell'}$ is the dilepton mass
of the two leptons from the Z boson decay prior to final state photon
radiation.
Only events decaying into the respective final state are considered in both the
numerator and denominator of this fraction.
The resulting cross section values are reported
in Table~\ref{tab:xsections} for the four leptonic channels.
There is good agreement among the four channels for both the 7 and 8\TeV data.
\begin{table*}[htb]
\centering
\topcaption{Measured WZ cross section in the four leptonic channels
at $\sqrt{s} = 7$ and 8\TeV.\label{tab:xsections}}
\begin{tabular}{lc}
\hline
Channel     & $\sigma({\Pp\Pp}\to{\PW\Z};\ \sqrt{s} = 7\TeV)$\,[pb] \\ \hline
$\Pe\Pe\Pe$             & $22.46 \pm  3.12\stat
\pm 0.43\thy \pm 1.33\exper \pm 0.49\lum$ \\
$\Pe\Pe\mu$     & $19.04 \pm 2.75\stat \pm  0.36 ( \mathrm{theo}) \pm 1.50\exper
\pm 0.42\lum$ \\
$\mu\mu\Pe$   & $19.13 \pm 2.60\stat \pm 0.37\thy \pm 1.56\exper
\pm 0.42\lum$ \\
$\mu\mu\mu$ & $20.36 \pm 2.31\stat \pm  0.39\thy \pm 1.48\exper
 \pm 0.45\lum$ \\ \hline
\\[2ex]
\hline
Channel     & $\sigma({\Pp\Pp}\to{\PW\Z};\ \sqrt{s} = 8\TeV)$\,[pb] \\ \hline
eee         & $24.80 \pm 1.92\stat
\pm  0.82 ( \mathrm{theo}) \pm 1.53\exper
\pm 0.64\lum$ \\
$\Pe\Pe\mu$     & $22.38 \pm 1.62\stat
\pm  0.74 ( \mathrm{theo}) \pm 1.78 \exper
\pm 0.58\lum$ \\
$\mu\mu\Pe$  & $23.94 \pm 1.52\stat
\pm 0.79  ( \mathrm{theo}) \pm 1.66\exper
\pm 0.62\lum$ \\
$\mu\mu\mu$ & $24.93 \pm 1.29\stat
\pm 0.83  ( \mathrm{theo}) \pm 2.14 \exper
\pm 0.65\lum$ \\ \hline
\end{tabular}

\end{table*}

\begin{figure}[hbt]
  \centering
    \includegraphics[width=\cmsFigWidth]{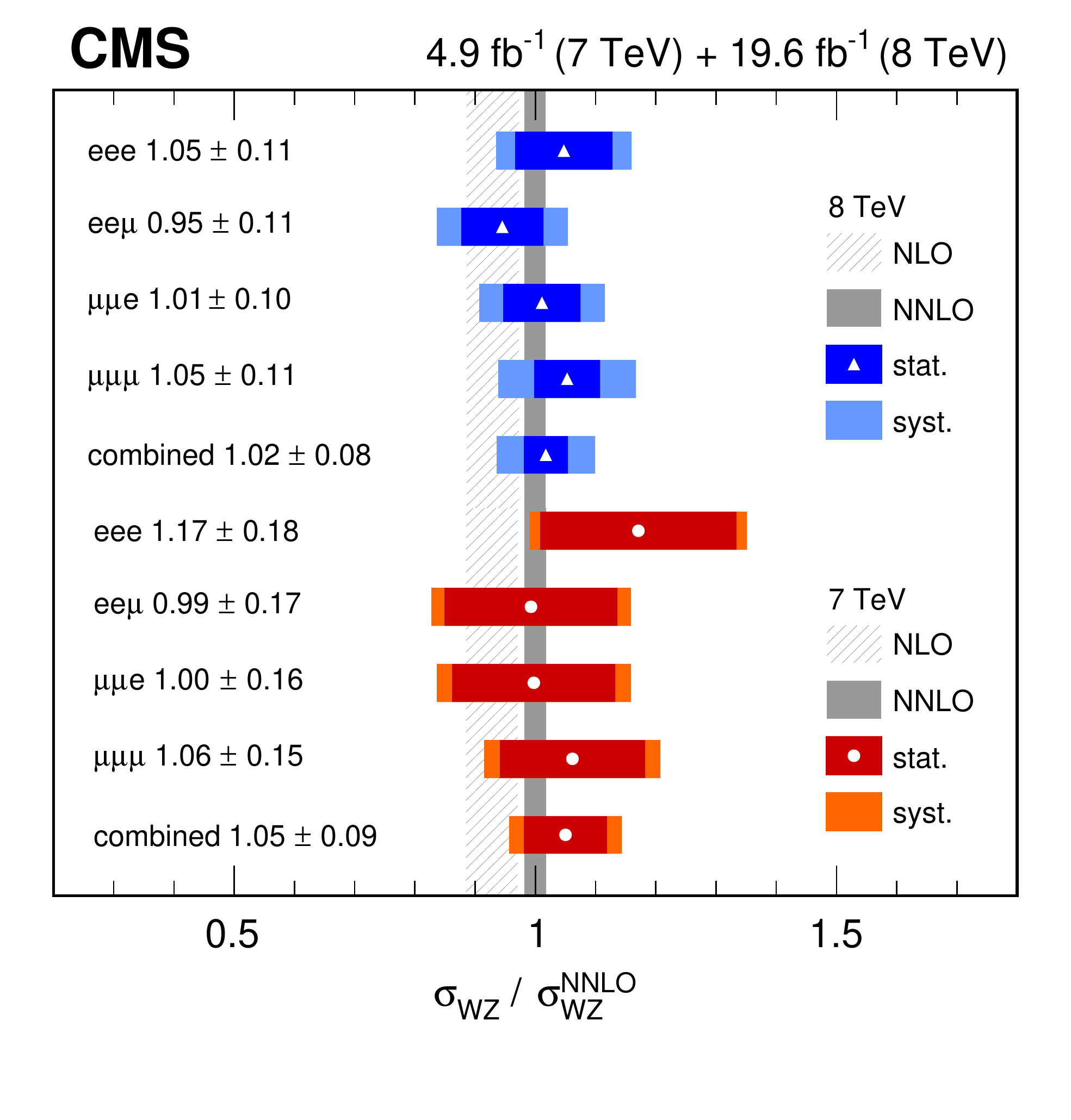}
    \caption{Ratio of measured inclusive cross sections to
    NNLO predictions. The vertical gray bands represent the theoretical
    uncertainties at 7 and 8\TeV.\label{fig:xsr}}
\end{figure}

These four measurements are combined
using the best linear unbiased estimator (BLUE)
method~\cite{Lyons:1988rp}. We have assumed full correlation for
all uncertainties common to different channels.
Combining the four leptonic channels, the total WZ cross section for
$71 < m_{\Z} < 111\GeV$, at 7 and 8\TeV, is measured to be
\ifthenelse{\boolean{cms@external}}{
\begin{multline*}
\sigma({\Pp\Pp}\to{\PW\Z};~\sqrt{s} = 7\TeV) = \\ \quad 20.14 \pm 1.32\stat
   \pm  0.38\thy\pm 1.06\exper  \pm 0.44\lum\unit{pb}.\\
\shoveleft{\sigma({\Pp\Pp}\to{\PW\Z};~\sqrt{s} = 8\TeV) =}\\ \quad 24.09 \pm 0.87\stat
 \pm  0.80\thy\pm 1.40\exper \pm 0.63\lum\unit{pb}.
\end{multline*}
}{
\begin{equation*}
\begin{aligned}
\sigma({\Pp\Pp}\to{\PW\Z};~\sqrt{s} = 7\TeV) &= 20.14 \pm 1.32\stat \pm  0.38\thy\pm 1.06\exper  \pm 0.44\lum\unit{pb}. \\
\sigma({\Pp\Pp}\to{\PW\Z};~\sqrt{s} = 8\TeV) &= 24.09 \pm 0.87\stat \pm  0.80\thy\pm 1.40\exper \pm 0.63\lum\unit{pb}.
\end{aligned}
\end{equation*}
}
{\tolerance=1200
These results can be compared with recent calculations
at NLO and next-to-next-to-leading order (NNLO)
in QCD via \textsc{Matrix}~\cite{Grazzini:2016swo}.
The NLO (NNLO) predictions are  $17.72^{+5.3\%}_{-1.8\%}$
($19.18_{-1.8\%}^{+1.7\%}$)\unit{pb}
at 7\TeV, and $21.80^{+5.1\%}_{-3.9\%}$ ($23.68 \pm 1.8\%$)\unit{pb}
at 8\TeV, where uncertainties include only scale variations.
All these predictions are in agreement with the measured values
within uncertainties.
The NLO predictions are slightly lower than the measured values,
and a better agreement is observed for the NNLO observations at
both centre-of-mass energies.
The ratios of the inclusive cross sections for the individual and combined results
to the NLO and NNLO predictions are shown in Fig.~\ref{fig:xsr}.
\par}

\begin{figure}[hbt]
  \centering
    \includegraphics[width=\cmsFigWidth]{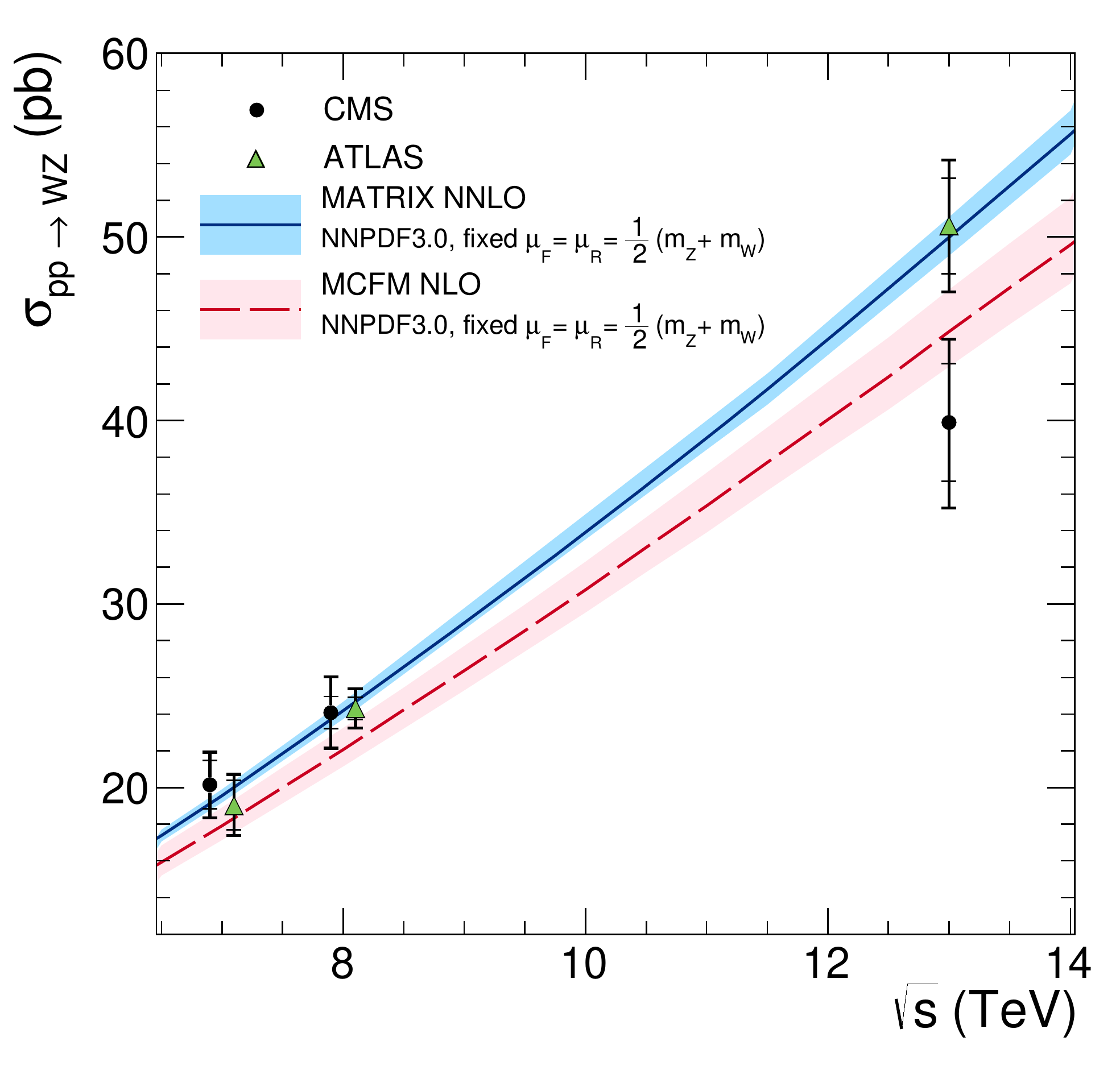}
    \caption{
    The WZ total cross section as a function of the proton-proton centre-of-mass energy. Results
    from the CMS and ATLAS experiments are compared to the predictions of MCFM and \textsc{Matrix}. The
    data uncertainties are statistical (inner bars) and statistical plus systematic added
    in quadrature (outer bars).
    The uncertainties covered by the band around the theoretical predictions are described in the text.
    The theoretical predictions
    and the CMS 13\TeV cross section are calculated for the Z boson mass window 60--120\GeV. The CMS
    7 and 8\TeV cross sections presented in this paper are calculated for the Z boson mass window 71--111\GeV
    (estimated correction factor 2\%), while all ATLAS measurements are performed with the Z boson
    mass window 66--116\GeV (1\%).
    \label{fig:wzxs_vs_sqrts}}
\end{figure}

The total WZ production cross sections for different centre-of-mass energies from the CMS~\cite{Khachatryan:2016tgp}
and ATLAS~\cite{Aad:2012twa,Aad:2016ett,Aaboud:2016yus} experiments are compared to theoretical predictions
calculated with MCFM~(NLO) and \textsc{Matrix}~(NNLO) in Fig.~\ref{fig:wzxs_vs_sqrts}. The theoretical predictions
describe, within the uncertainties, the energy dependence of the measured cross sections.
The band around the theoretical predictions in this figure reflects uncertainties generated by
varying the factorization and renormalization scales up and down by a factor of two and
also the (PDF+$\alpha_\mathrm{S}$) uncertainty of NNPDF3.0 for NLO predictions.

\subsection{Differential cross section measurement}

Using the larger available integrated luminosity in the 8\TeV sample, we
measure the
differential WZ cross sections as a function of three different observables: the
Z boson \pt, the number of jets produced in
association with the
$\ell \nu \ell'\ell'$ final state, and the \pt of the
leading accompanying jet. For the latter two measurements, the differential
cross sections are defined for generated jets built from all stable particles
using the anti-\kt algorithm~\cite{Cacciari:2008gp} with a distance
parameter of 0.5,
but excluding the electrons, muons, and neutrinos from the $\PW$ and $\cPZ$ boson decays.
Jets are required to have $\pt > 30\GeV$ and $\abs{\eta}<2.5$. They also
must be separated from the charged leptons from the W and Z boson decays by
$\Delta R(\text{jet},\ell)>0.5$. The jets reconstructed from PF
candidates, clustered by the same algorithm, have to fulfill the same
requirements.

To obtain the cross section in each bin, the background contribution
is first subtracted from the observed yield in each bin, in the same way
as it was done for the inclusive cross section.
The measured signal spectra are then  corrected for the detector
effects. These include efficiencies as well as bin-to-bin migrations
due to finite resolution. Both effects
are  treated using the iterative D'Agostini unfolding
technique~\cite{D'Agostini:1994zf},  as implemented
in \textsc{RooUnfold}~\cite{RooUnfold},  with 5 iterations.
The technique uses response matrices that
relate the true distribution of an
observable to the observed distribution after
including detector effects.
The response matrices are obtained using the
signal MC sample for all four leptonic final states separately.
The unfolded spectra are then used to obtain differential cross
sections for all four leptonic final states. The four channels are
combined bin-by-bin.

\begin{figure}[htp]
\centering
   \includegraphics[scale=0.4]{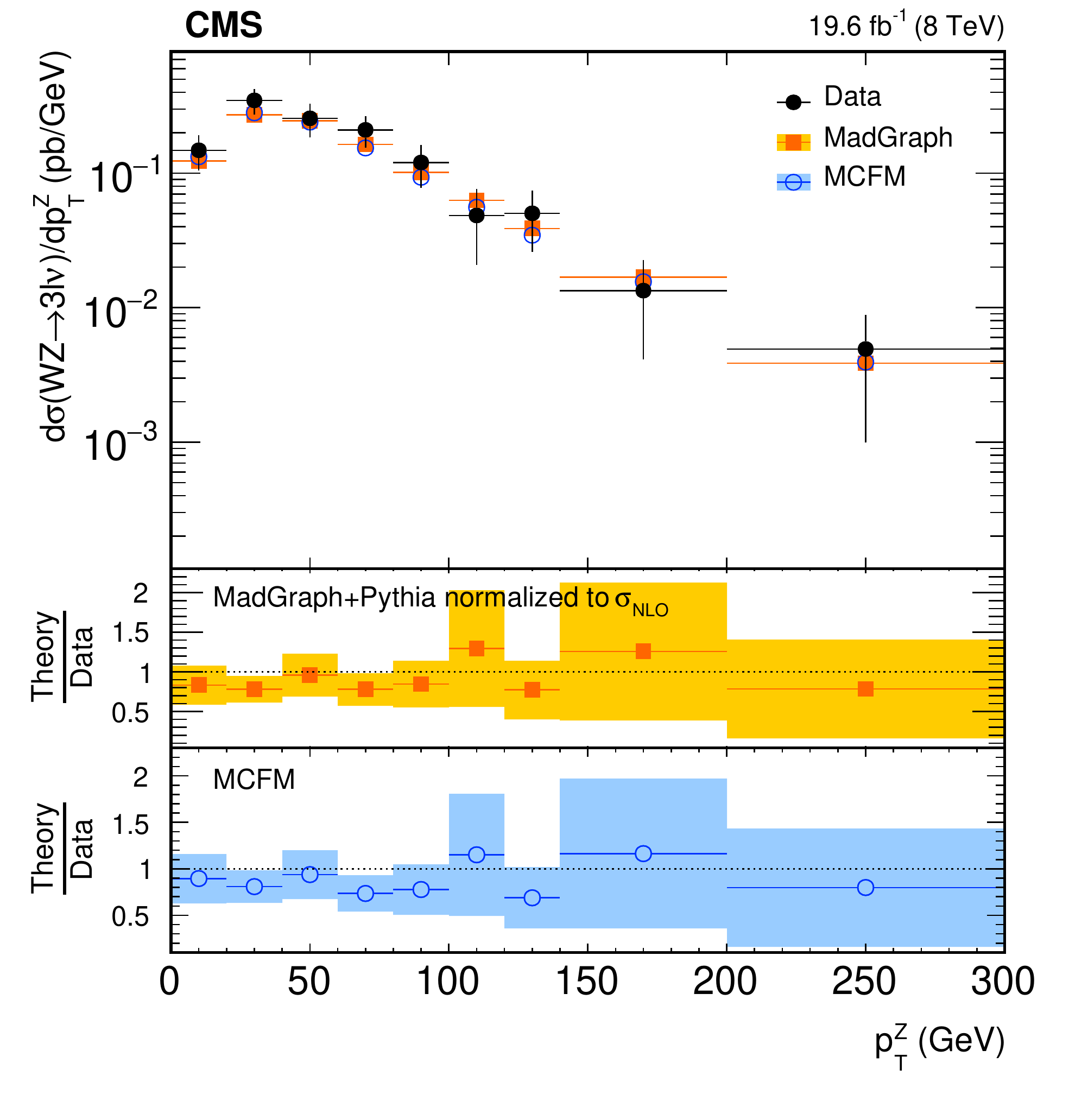}
\caption{Differential WZ cross section at $\sqrt{s}=8\TeV$
as a function of the Z boson transverse momentum.
The measurement is compared with \MCFM and \MADGRAPH predictions.
The  \MADGRAPH prediction is rescaled to the total NLO cross
section as predicted by \MCFM. The error bands in the ratio plots
indicate the relative errors on the data in each bin and contain both
statistical and systematic uncertainties.
}
\label{fig:diffxs_zpt}
\end{figure}

\begin{figure}[htp]
\centering

   \includegraphics[width=0.45\textwidth]{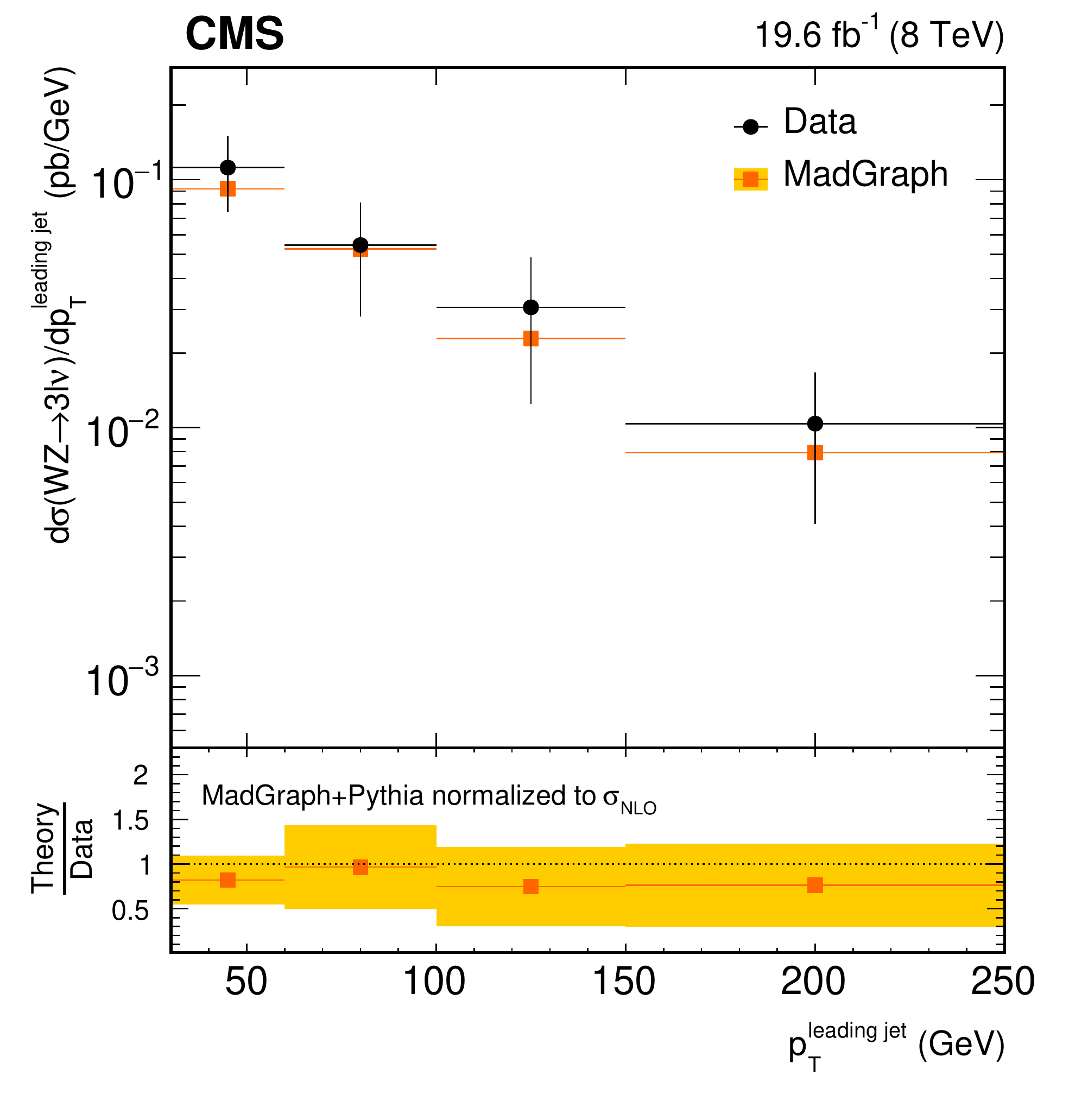}
   \includegraphics[width=0.45\textwidth]{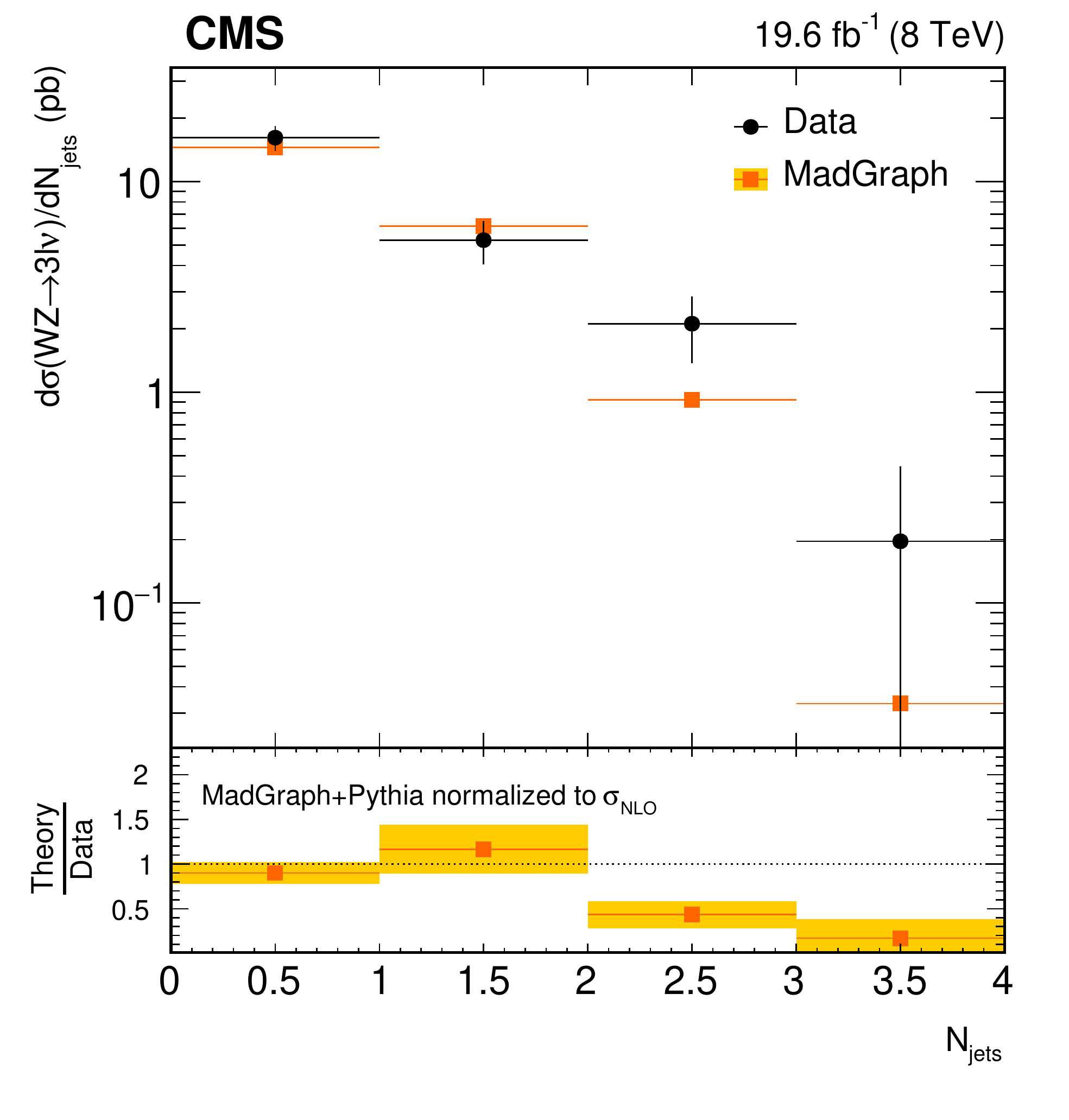}
\caption{Differential WZ cross section at $\sqrt{s}=8\TeV$
as a function of:  (\cmsLeft) the leading jet
transverse momentum; (\cmsRight) the number of accompanying jets.
The measurements are compared with \MADGRAPH predictions.
The \MADGRAPH prediction is rescaled to the total NLO cross
section as predicted by \MCFM. The error bands in the ratio plots
indicate the relative errors on the data in each bin and contain both
statistical and systematic uncertainties.
}
\label{fig:diffxs_jets}
\end{figure}

A few additional sources of systematic uncertainties need to be
considered with respect to those described in Section~\ref{sec:systematics}.
The measurements involving jets are affected by the experimental
uncertainties in the jet energy scale and resolution. The effects on
the response matrices are studied by smearing and scaling the
jet energies within their uncertainties. Furthermore, an uncertainty
due to the limited size of the simulated sample used to build the
response matrices is also included. The unfolding procedure
introduces statistical correlations between bins, which range from a few percent
up to 40\% in a few cases. These correlations are taken into account together
with correlated systematic uncertainties
by using a generalization of the BLUE method as described in Ref.~\cite{Valassi:2003mu}.
The three measured differential cross sections are given in
Tables~\ref{tab:xsdiff_channels_Zpt}, \ref{tab:xsdiff_channels_Njets},
and~\ref{tab:xsdiff_channels_LeadingJetPt} for each of the four final
states,
and the combined results are given in Table~\ref{tab:xsdiff_combined}.
The combined differential cross sections are shown in
Figs.~\ref{fig:diffxs_zpt} and \ref{fig:diffxs_jets}.

\begin{table*}[htbp]
\centering
\topcaption{Differential WZ cross section as a function of the
Z transverse momentum at $\sqrt{s}=8\TeV$ for the
four leptonic final states.
The first uncertainty is statistical, the second is systematic, and the
third is the integrated luminosity. \label{tab:xsdiff_channels_Zpt}
}
\begin{tabular} { l|c|c|c|c}
\hline
$\pt^{\cPZ}$          & \multicolumn{4}{c}{ $\rd\sigma / \rd \pt^{\cPZ}$\,[pb /\GeVns{}]}
   \\ \cline{2-5}
   [\GeVns{}] & $\Pe\Pe\Pe$ & $\Pe\Pe\mu$ & $\mu\mu\Pe$ & $\mu\mu\mu$ \\
\hline
0--20&
\parbox{0.2\textwidth}
 { \(
 \begin{array}{lll}
(1.63 & \pm &   0.90  \\
 & \pm &   0.22  \\
 & \pm &   0.04)\\
 & & \times 10 ^{-1}
\end{array}
\) }
&
\parbox{0.2\textwidth}
 { \(
 \begin{array}{lll}
(9.3 & \pm &6.8  \\
 & \pm &  1.3  \\
 & \pm &  0.2)\\
 & & \times 10 ^{-2}
\end{array}
\) }
&
\parbox{0.2\textwidth}
 { \(
 \begin{array}{lll}
(1.68 & \pm &  0.92  \\
 & \pm &  0.21  \\
 & \pm &  0.04)\\
 & & \times 10 ^{-1}
\end{array}
\) }
&
\parbox{0.2\textwidth}
 { \(
 \begin{array}{lll}
(2.01 & \pm &  1.00  \\
 & \pm &  0.20  \\
 & \pm &  0.05)\\
  & & \times 10 ^{-1}
\end{array}
\) }

\\ \hline

20--40&
\parbox{0.2\textwidth}
 { \(
 \begin{array}{lll}
(3.9 & \pm &   1.4  \\
 & \pm &   0.5  \\
 & \pm &   0.1)\\
 & & \times 10 ^{-1}
\end{array}
\) }
&
\parbox{0.2\textwidth}
 { \(
 \begin{array}{lll}
(3.17 & \pm &1.26  \\
 & \pm &  0.39  \\
 & \pm &  0.08)\\
 & & \times 10 ^{-1}
\end{array}
\) }
&
\parbox{0.2\textwidth}
 { \(
 \begin{array}{lll}
(2.76 & \pm &  1.18  \\
 & \pm &  0.62  \\
 & \pm &  0.07)\\
 & & \times 10 ^{-1}
\end{array}
\) }
&
\parbox{0.2\textwidth}
 { \(
 \begin{array}{lll}
(3.42 & \pm &  1.31  \\
 & \pm &  0.57  \\
 & \pm &  0.09)\\
  & & \times 10 ^{-1}
\end{array}
\) }

\\ \hline

40--60&
\parbox{0.2\textwidth}
 { \(
 \begin{array}{lll}
(3.14 & \pm &   1.25  \\
 & \pm &   0.60  \\
 & \pm &   0.08)\\
 & & \times 10 ^{-1}
\end{array}
\) }
&
\parbox{0.2\textwidth}
 { \(
 \begin{array}{lll}
(2.70 & \pm &1.16  \\
 & \pm &  0.43  \\
 & \pm &  0.07)\\
 & & \times 10 ^{-1}
\end{array}
\) }
&
\parbox{0.2\textwidth}
 { \(
 \begin{array}{lll}
(2.29 & \pm &  1.07  \\
 & \pm &  0.48  \\
 & \pm &  0.06)\\
 & & \times 10 ^{-1}
\end{array}
\) }
&
\parbox{0.2\textwidth}
 { \(
 \begin{array}{lll}
(2.82 & \pm &  1.19  \\
 & \pm &  0.56  \\
 & \pm &  0.07)\\
  & & \times 10 ^{-1}
\end{array}
\) }

\\ \hline

60--80&
\parbox{0.2\textwidth}
 { \(
 \begin{array}{lll}
(1.69 & \pm &   0.92  \\
 & \pm &   0.30  \\
 & \pm &   0.04)\\
 & & \times 10 ^{-1}
\end{array}
\) }
&
\parbox{0.2\textwidth}
 { \(
 \begin{array}{lll}
(2.07 & \pm &1.02  \\
 & \pm &  0.31  \\
 & \pm &  0.05)\\
 & & \times 10 ^{-1}
\end{array}
\) }
&
\parbox{0.2\textwidth}
 { \(
 \begin{array}{lll}
(2.31 & \pm &  1.07  \\
 & \pm &  0.33  \\
 & \pm &  0.06)\\
 & & \times 10 ^{-1}
\end{array}
\) }
&
\parbox{0.2\textwidth}
 { \(
 \begin{array}{lll}
(2.03 & \pm &  1.01  \\
 & \pm &  0.31  \\
 & \pm &  0.05)\\
  & & \times 10 ^{-1}
\end{array}
\) }

\\ \hline

80--100&
\parbox{0.2\textwidth}
 { \(
 \begin{array}{lll}
(1.27 & \pm &   0.80  \\
 & \pm &   0.23  \\
 & \pm &   0.03)\\
 & & \times 10 ^{-1}
\end{array}
\) }
&
\parbox{0.2\textwidth}
 { \(
 \begin{array}{lll}
(1.02 & \pm &0.71  \\
 & \pm &  0.17  \\
 & \pm &  0.03)\\
 & & \times 10 ^{-1}
\end{array}
\) }
&
\parbox{0.2\textwidth}
 { \(
 \begin{array}{lll}
(1.30 & \pm &  0.81  \\
 & \pm &  0.25  \\
 & \pm &  0.03)\\
 & & \times 10 ^{-1}
\end{array}
\) }
&
\parbox{0.2\textwidth}
 { \(
 \begin{array}{lll}
(1.25 & \pm &  0.79  \\
 & \pm &  0.21  \\
 & \pm &  0.03)\\
  & & \times 10 ^{-1}
\end{array}
\) }

\\ \hline

100--120&
\parbox{0.2\textwidth}
 { \(
 \begin{array}{lll}
(8.1 & \pm &   6.4  \\
 & \pm &   2.2  \\
 & \pm &   0.2)\\
 & & \times 10 ^{-2}
\end{array}
\) }
&
\parbox{0.2\textwidth}
 { \(
 \begin{array}{lll}
(2.76 & \pm &3.72  \\
 & \pm &  1.55  \\
 & \pm &  0.07)\\
 & & \times 10 ^{-2}
\end{array}
\) }
&
\parbox{0.2\textwidth}
 { \(
 \begin{array}{lll}
(5.0 & \pm &  5.0  \\
 & \pm &  1.4  \\
 & \pm &  0.1)\\
 & & \times 10 ^{-2}
\end{array}
\) }
&
\parbox{0.2\textwidth}
 { \(
 \begin{array}{lll}
(7.8 & \pm &  6.3  \\
 & \pm &  1.4  \\
 & \pm &  0.2)\\
  & & \times 10 ^{-2}
\end{array}
\) }

\\ \hline

120--140&
\parbox{0.2\textwidth}
 { \(
 \begin{array}{lll}
(5.8 & \pm &   5.4  \\
 & \pm &   0.9  \\
 & \pm &   0.1)\\
 & & \times 10 ^{-2}
\end{array}
\) }
&
\parbox{0.2\textwidth}
 { \(
 \begin{array}{lll}
(6.2 & \pm &5.6  \\
 & \pm &  0.8  \\
 & \pm &  0.2)\\
 & & \times 10 ^{-2}
\end{array}
\) }
&
\parbox{0.2\textwidth}
 { \(
 \begin{array}{lll}
(3.12 & \pm &  3.95  \\
 & \pm &  1.13  \\
 & \pm &  0.08)\\
 & & \times 10 ^{-2}
\end{array}
\) }
&
\parbox{0.2\textwidth}
 { \(
 \begin{array}{lll}
(4.1 & \pm &  4.5  \\
 & \pm &  1.2  \\
 & \pm &  0.1)\\
  & & \times 10 ^{-2}
\end{array}
\) }

\\ \hline

140--200&
\parbox{0.2\textwidth}
 { \(
 \begin{array}{lll}
(1.07 & \pm &   1.34  \\
 & \pm &   0.58  \\
 & \pm &   0.03)\\
 & & \times 10 ^{-2}
\end{array}
\) }
&
\parbox{0.2\textwidth}
 { \(
 \begin{array}{lll}
(1.09 & \pm &1.35  \\
 & \pm &  0.62  \\
 & \pm &  0.03)\\
 & & \times 10 ^{-2}
\end{array}
\) }
&
\parbox{0.2\textwidth}
 { \(
 \begin{array}{lll}
(2.73 & \pm &  2.13  \\
 & \pm &  0.56  \\
 & \pm &  0.07)\\
 & & \times 10 ^{-2}
\end{array}
\) }
&
\parbox{0.2\textwidth}
 { \(
 \begin{array}{lll}
(1.46 & \pm &  1.56  \\
 & \pm &  0.53  \\
 & \pm &  0.04)\\
  & & \times 10 ^{-2}
\end{array}
\) }

\\ \hline

200--300&
\parbox{0.2\textwidth}
 { \(
 \begin{array}{lll}
(3.66 & \pm &   6.05  \\
 & \pm &   1.58  \\
 & \pm &   0.10)\\
 & & \times 10 ^{-3}
\end{array}
\) }
&
\parbox{0.2\textwidth}
 { \(
 \begin{array}{lll}
(9.0 & \pm &9.5  \\
 & \pm &  1.7  \\
 & \pm &  0.2)\\
 & & \times 10 ^{-3}
\end{array}
\) }
&
\parbox{0.2\textwidth}
 { \(
 \begin{array}{lll}
(7.4 & \pm &  8.6  \\
 & \pm &  1.7  \\
 & \pm &  0.2)\\
 & & \times 10 ^{-3}
\end{array}
\) }
&
\parbox{0.2\textwidth}
 { \(
 \begin{array}{lll}
(5.8 & \pm &  7.6  \\
 & \pm &  1.8  \\
 & \pm &  0.2)\\
  & & \times 10 ^{-3}
\end{array}
\) }
\\ \hline
\end{tabular}
\end{table*}

\begin{table*}[htbp]
\centering
\topcaption{Differential WZ cross section as a function of the jet multiplicity
at $\sqrt{s}=8\TeV$ for  the four leptonic final
states.
 Notations are as in Table~\ref{tab:xsdiff_channels_Zpt}.
 \label{tab:xsdiff_channels_Njets}
}
\begin{tabular} { l|c|c|c|c}
\hline

$N_\text{jets}$          & \multicolumn{4}{c}{ $\rd\sigma / \rd N_\text{jets} $\,[pb]}
   \\ \cline{2-5}
            & $\Pe\Pe\Pe$ & $\Pe\Pe\mu$ & $\mu\mu\Pe$ & $\mu\mu\mu$ \\
\hline

0 jets&
\parbox{0.2\textwidth}
 { \(
 \begin{array}{lll}
16.60 & \pm &   4.07  \\
 & \pm &   1.04  \\
 & \pm &   0.43
\end{array}
\) }
&
\parbox{0.2\textwidth}
 { \(
 \begin{array}{lll}
15.68 & \pm &   3.96  \\
 & \pm &   1.03  \\
 & \pm &   0.41
\end{array}
\) }
&
\parbox{0.2\textwidth}
 { \(
 \begin{array}{lll}
14.97 & \pm &  3.87  \\
 & \pm &  0.93  \\
 & \pm &  0.39
\end{array}
\) }
&
\parbox{0.2\textwidth}
 { \(
 \begin{array}{lll}
18.78 & \pm &  4.33 \\
 & \pm &  1.11\\
 & \pm &  0.49
\end{array}
\) }

\\ \hline

1 jet&
\parbox{0.2\textwidth}
 { \(
 \begin{array}{lll}
6.06 & \pm &   2.46  \\
 & \pm &   0.48  \\
 & \pm &   0.16
\end{array}
\) }
&
\parbox{0.2\textwidth}
 { \(
 \begin{array}{lll}
4.80 & \pm &   2.19  \\
 & \pm &   0.57  \\
 & \pm &   0.12
\end{array}
\) }
&
\parbox{0.2\textwidth}
 { \(
 \begin{array}{lll}
5.32 & \pm &  2.31  \\
 & \pm &  0.61  \\
 & \pm &  0.14
\end{array}
\) }
&
\parbox{0.2\textwidth}
 { \(
 \begin{array}{lll}
4.84 & \pm &  2.20 \\
 & \pm &  0.72\\
 & \pm &  0.13
\end{array}
\) }

\\ \hline

2 jets&
\parbox{0.2\textwidth}
 { \(
 \begin{array}{lll}
2.43 & \pm &   1.56  \\
 & \pm &   0.34  \\
 & \pm &   0.06
\end{array}
\) }
&
\parbox{0.2\textwidth}
 { \(
 \begin{array}{lll}
1.75 & \pm &   1.32  \\
 & \pm &   0.32  \\
 & \pm &   0.05
\end{array}
\) }
&
\parbox{0.2\textwidth}
 { \(
 \begin{array}{lll}
2.93 & \pm &  1.71  \\
 & \pm &  0.26  \\
 & \pm &  0.08
\end{array}
\) }
&
\parbox{0.2\textwidth}
 { \(
 \begin{array}{lll}
1.54 & \pm &  1.24 \\
 & \pm &  0.32\\
 & \pm &  0.04
\end{array}
\) }

\\ \hline

3 jets&
\parbox{0.2\textwidth}
 { \(
 \begin{array}{lll}
(7.8 & \pm &   27.9  \\
 & \pm &   7.3  \\
 & \pm &   0.2)\\
 & & \times 10 ^{-2}
\end{array}
\) }
&
\parbox{0.2\textwidth}
 { \(
 \begin{array}{lll}
0.45 & \pm &   0.67  \\
 & \pm &   0.17  \\
 & \pm &   0.01
\end{array}
\) }
&
\parbox{0.2\textwidth}
 { \(
 \begin{array}{lll}
0.42 & \pm &  0.65  \\
 & \pm &  0.21  \\
 & \pm &  0.01
\end{array}
\) }
&
\parbox{0.2\textwidth}
 { \(
 \begin{array}{lll}
0.79 & \pm &  0.89 \\
 & \pm &  0.26\\
 & \pm &  0.02
\end{array}
\) }
\\ \hline
\end{tabular}
\end{table*}

\begin{table*}[htbp]
\centering
\topcaption{Differential WZ cross section as a function of the
leading jet transverse momentum at $\sqrt{s}=8\TeV$ for the
four leptonic final states. Notations are as in
Table~\ref{tab:xsdiff_channels_Zpt}. \label{tab:xsdiff_channels_LeadingJetPt}
}

\begin{tabular} { l|c|c|c|c}
\hline
$\pt^{\text{leading jet}}$          & \multicolumn{4}{c}{ $\rd\sigma / \rd \pt^\text{leading jet} $\,[pb/\GeVns{}]}
   \\ \cline{2-5}
 [\GeVns{}] & $\Pe\Pe\Pe$ & $\Pe\Pe\mu$ & $\mu\mu\Pe$ & $\mu\mu\mu$ \\
\hline

30--60&
\parbox{0.2\textwidth}
 { \(
 \begin{array}{lll}
(1.22 & \pm &   0.64  \\
 & \pm &   0.34  \\
 & \pm &   0.03)\\
 & & \times 10 ^{-1}
\end{array}
\) }
&
\parbox{0.2\textwidth}
 { \(
 \begin{array}{lll}
(1.11 & \pm &0.61  \\
 & \pm &  0.20  \\
 & \pm &  0.03)\\
 & & \times 10 ^{-1}
\end{array}
\) }
&
\parbox{0.2\textwidth}
 { \(
 \begin{array}{lll}
(1.10 & \pm &  0.61  \\
 & \pm &  0.24  \\
 & \pm &  0.03)\\
 & & \times 10 ^{-1}
\end{array}
\) }
&
\parbox{0.2\textwidth}
 { \(
 \begin{array}{lll}
(1.02 & \pm &  0.58  \\
 & \pm &  0.24  \\
 & \pm &  0.03)\\
  & & \times 10 ^{-1}
\end{array}
\) }

\\ \hline

60--100&
\parbox{0.2\textwidth}
 { \(
 \begin{array}{lll}
(5.4 & \pm &   3.7  \\
 & \pm &   1.7  \\
 & \pm &   0.1)\\
 & & \times 10 ^{-2}
\end{array}
\) }
&
\parbox{0.2\textwidth}
 { \(
 \begin{array}{lll}
(4.3 & \pm &3.3  \\
 & \pm &  2.1  \\
 & \pm &  0.1)\\
 & & \times 10 ^{-2}
\end{array}
\) }
&
\parbox{0.2\textwidth}
 { \(
 \begin{array}{lll}
(6.5 & \pm &  4.0  \\
 & \pm &  2.0  \\
 & \pm &  0.2)\\
 & & \times 10 ^{-2}
\end{array}
\) }
&
\parbox{0.2\textwidth}
 { \(
 \begin{array}{lll}
(6.3 & \pm &  4.0  \\
 & \pm &  2.3  \\
 & \pm &  0.2)\\
  & & \times 10 ^{-2}
\end{array}
\) }

\\ \hline

100--150&
\parbox{0.2\textwidth}
 { \(
 \begin{array}{lll}
(2.96 & \pm &   2.43  \\
 & \pm &   1.57  \\
 & \pm &   0.08)\\
 & & \times 10 ^{-2}
\end{array}
\) }
&
\parbox{0.2\textwidth}
 { \(
 \begin{array}{lll}
(3.26 & \pm &2.55  \\
 & \pm &  1.40  \\
 & \pm &  0.08)\\
 & & \times 10 ^{-2}
\end{array}
\) }
&
\parbox{0.2\textwidth}
 { \(
 \begin{array}{lll}
(3.9 & \pm &  2.8  \\
 & \pm &  1.2  \\
 & \pm &  0.1)\\
 & & \times 10 ^{-2}
\end{array}
\) }
&
\parbox{0.2\textwidth}
 { \(
 \begin{array}{lll}
(2.44 & \pm &  2.21  \\
 & \pm &  1.32  \\
 & \pm &  0.06)\\
  & & \times 10 ^{-2}
\end{array}
\) }

\\ \hline

150--250&
\parbox{0.2\textwidth}
 { \(
 \begin{array}{lll}
(1.18 & \pm &   1.09  \\
 & \pm &   0.29  \\
 & \pm &   0.03)\\
 & & \times 10 ^{-2}
\end{array}
\) }
&
\parbox{0.2\textwidth}
 { \(
 \begin{array}{lll}
(8.1 & \pm &9.0  \\
 & \pm &  3.4  \\
 & \pm &  0.2)\\
 & & \times 10 ^{-3}
\end{array}
\) }
&
\parbox{0.2\textwidth}
 { \(
 \begin{array}{lll}
(1.07 & \pm &  1.03  \\
 & \pm &  0.61  \\
 & \pm &  0.03)\\
 & & \times 10 ^{-2}
\end{array}
\) }
&
\parbox{0.2\textwidth}
 { \(
 \begin{array}{lll}
(1.00 & \pm &  1.00  \\
 & \pm &  0.42  \\
 & \pm &  0.03)\\
  & & \times 10 ^{-2}
\end{array}
\) }
\\
\hline
\end{tabular}
\end{table*}

\begin{table*}[htbp]
\centering
\topcaption{Combined result for the differential WZ cross sections at $\sqrt{s}=8\TeV$.
\label{tab:xsdiff_combined}}
\begin{tabular} { ll}
\hline
$\pt^{\cPZ}$ [\GeVns{}] & \multicolumn{1}{c}{$\rd\sigma / \rd \pt^{\cPZ}$  [pb/\GeVns{}] }\\
\hline
0--20    & [1.48 $\pm$ 0.40\stat $\pm$ 0.17\syst $\pm$ 0.04\lum]$\times 10 ^{-1}$ \\
20--40   & [3.47 $\pm$ 0.60\stat $\pm$ 0.50\syst $\pm$ 0.09\lum]$\times 10 ^{-1}$ \\
40--60   & [2.56 $\pm$ 0.54\stat $\pm$ 0.49\syst $\pm$ 0.07\lum]$\times 10 ^{-1}$ \\
60--80   & [2.10 $\pm$ 0.47\stat $\pm$ 0.30\syst $\pm$ 0.05\lum]$\times 10 ^{-1}$ \\
80--100  & [1.20 $\pm$ 0.37\stat $\pm$ 0.21\syst $\pm$ 0.03\lum]$\times 10 ^{-1}$ \\
100--120 & [4.9  $\pm$ 2.3 \stat $\pm$ 1.5 \syst $\pm$ 0.1 \lum]$\times 10 ^{-2}$ \\
120--140 & [5.0  $\pm$ 2.2 \stat $\pm$ 1.0 \syst $\pm$ 0.1 \lum]$\times 10 ^{-2}$ \\
140--200 & [1.34 $\pm$ 0.73\stat $\pm$ 0.57\syst $\pm$ 0.03\lum]$\times 10 ^{-2}$ \\
200--300 & [4.9  $\pm$ 3.6 \stat $\pm$ 1.6 \syst $\pm$ 0.1 \lum]$\times 10 ^{-3}$ \\
\hline
\\[2ex]\hline
$N_\text{jets}$ & \multicolumn{1}{c}{$\rd\sigma / \rd N_\text{jets} $  [pb]} \\
\hline
0 jets & 16.15 $\pm$ 1.95 \stat $\pm$ 0.88 \syst $\pm$ 0.42 \lum \\
1 jet  & 5.27  $\pm$ 1.11 \stat $\pm$ 0.52 \syst $\pm$ 0.14 \lum \\
2 jets & 2.11  $\pm$ 0.69 \stat $\pm$ 0.27 \syst $\pm$ 0.05 \lum \\
3 jets & 0.196 $\pm$ 0.227\stat $\pm$ 0.102\syst $\pm$ 0.005\lum \\
\hline\\[2ex]
\hline
$\pt^\text{leading jet}$ [\GeVns{}] & \multicolumn{1}{c}{$\rd\sigma / \rd \pt^\text{leading jet} $  [pb/\GeVns{}]} \\
\hline
30--60   & [1.12 $\pm$ 0.30\stat $\pm$ 0.23\syst $\pm$ 0.03\lum]$\times 10 ^{-1}$ \\
60--100  & [5.5  $\pm$ 1.8 \stat $\pm$ 1.9 \syst $\pm$ 0.1 \lum]$\times 10 ^{-2}$ \\
100--150 & [3.06 $\pm$ 1.20\stat $\pm$ 1.37\syst $\pm$ 0.08\lum]$\times 10 ^{-2}$ \\
150--250 & [1.04 $\pm$ 0.48\stat $\pm$ 0.41\syst $\pm$ 0.03\lum]$\times 10 ^{-2}$ \\
\hline
\end{tabular}
\end{table*}

The differential cross sections are compared with the \MCFM and
\MADGRAPH predictions. The \MADGRAPH
spectra are normalized to the NLO cross section as predicted
by MCFM.

\subsection{Anomalous triple gauge couplings limits}

Triple gauge boson couplings are a consequence of the non-Abelian nature of the
SM electroweak sector. Several extensions of the SM predict additional processes
with multiple bosons in the final state so any observed
deviation of diboson production cross sections from their SM predictions
could be an early sign of new physics. The most general Lorentz invariant
effective Lagrangian that describes WWV couplings, where $\PV =
\gamma$ or Z,  has 14 independent
parameters~\cite{hagiwara1,hagiwara2},
seven for $\PV=\gamma $  and seven for $\PV = \cPZ$. Assuming
charge conjugation (C) and parity (P)  conservation, only six independent parameters
remain. The effective Lagrangian, normalized by the electroweak coupling, is
given by:
\ifthenelse{\boolean{cms@external}}{
\begin{multline}
\label{eq:atgc_lagrangian}
\frac{\mathcal{L}_\mathrm{TGC}}{g_{\PW\PW\PV}} =
  ig_1^{\PV}(W^{-}_{\mu\nu} {W}^{+\mu} {V}^\nu - {W}^{-}_\mu {V}_\nu {W}^{+\mu\nu})\\
 + i\kappa_{\PV} {W}^{-}_\mu {W}^{+}_\nu {V}^{\mu\nu} +
   \frac{i\lambda_{\PV}}{M_{\PW}^2}W^{-}_{\delta\mu}{W}^{+\mu}_\nu {V}^{\nu\delta},
\end{multline}
}{
\begin{equation}
\label{eq:atgc_lagrangian}
\frac{\mathcal{L}_\mathrm{TGC}}{g_{\PW\PW\PV}} =
  ig_1^{\PV}(W^{-}_{\mu\nu} {W}^{+\mu} {V}^\nu - {W}^{-}_\mu {V}_\nu {W}^{+\mu\nu})
 + i\kappa_{\PV} {W}^{-}_\mu {W}^{+}_\nu {V}^{\mu\nu} +
   \frac{i\lambda_{\PV}}{M_{\PW}^2}W^{-}_{\delta\mu}{W}^{+\mu}_\nu {V}^{\nu\delta},
\end{equation}
}
where
${{W^\pm}}_{\mu\nu} = \partial_\mu {W}^\pm_\nu - \partial_\nu {W}^\pm_\mu$,
${V}_{\mu\nu} = \partial_\mu V_\nu - \partial_\nu {V}_\mu$,
and couplings $g_{\PW\PW\gamma} = -e$ and $g_{\PW\PW\Z} = -e \cot\theta_{\PW}$,
with $\theta_{\PW}$ being the weak mixing angle. Assuming
electromagnetic gauge invariance, \ie $g_1^\gamma = 1$, the remaining
parameters that describe the WWV coupling are
$g_1^{\Z}$, $\kappa_{\Z}$, $\kappa_\gamma$, $\lambda_{\Z}$ and
$\lambda_\gamma$. In the SM $\lambda_{\Z} = \lambda_\gamma = 0$ and
$g_1^{\Z} = \kappa_{\Z} = \kappa_\gamma = 1$. The couplings are further
reduced to three independent parameters if one requires the Lagrangian to be
${\mathcal{SU}}(2)_L\times {\mathcal{U}}(1)_Y$ invariant (``LEP parameterization'')
\cite{GrosseKnetter:1993rp,Bilenky:1993ms,Bilenky:1993uy}:
\begin{equation}
\label{eq:hisz}
\Delta\kappa_{\Z} = \Delta g_1^{\Z} - \Delta \kappa_\gamma \, \tan^2\theta_{\PW},\quad
\lambda = \lambda_\gamma = \lambda_{\Z},
\end{equation}

where $\Delta\kappa_{\Z}=\kappa_{\Z}-1$, $\Delta g_1^{\Z}= g_1^{\Z}-1$ and $\Delta \kappa_\gamma=\kappa_\gamma-1$.

In this analysis we measure $\Delta\kappa_{\Z}$, $\lambda$, and
$\Delta g_1^{\Z}$ from WZ production at 8\TeV.  No form factor
scaling is used for aTGCs, as this allows
us to provide results without the bias that can be caused by the choice of the
form factor energy dependence.

Another approach to the parametrization of anomalous couplings is through effective field
theory (EFT), with the higher-order operators added to the SM Lagrangian as follows:
\begin{equation}
\label{eq:eft_lagrangian}
\mathcal{L}_\mathrm{EFT} = \mathcal{L}_\mathrm{ SM}+\sum_{n=1}^{\infty}\sum_{i}\frac{c_{i}^{(n)}}{\Lambda^{n}}O_{i}^{(n+4)}.
\end{equation}
Here $O_{i}$ are the higher-order operators, the coefficients $c_{i}$ are dimensionless,
and $\Lambda$ is the mass scale of new physics. Operators are suppressed if the accessible energy
is low compared to the mass scale. There are three CP-even operators that
contribute to WWZ TGC, $O_{\PW\PW\PW}$, $O_{\PW}$, and $O_{\PB}$. For the case
of `LEP parametrization' and no form factor scaling of aTGCs,
the relations between parameters in the aTGCs and EFT approaches are as follows:
\begin{equation*}
\begin{aligned}
\label{eq:eft_to_aTGC}
g_{1}^{\Z} &= 1+c_{\PW}\frac{m_{\Z}^2}{2\Lambda^2},\\
\kappa_{\gamma} &= 1+\left(c_{\PW}+c_{\PB}\right)\frac{m_{\PW}^2}{2\Lambda^2},\\
\kappa_{\Z} &= 1+\left(c_{\PW}-c_{\PB} \tan^2\theta_{\PW} \right)\frac{m_{\PW}^2}{2\Lambda^2},\\
\lambda_{\Z} &= \lambda_{\gamma}  = c_{\PW\PW\PW}\frac{3 g^2 m_{\PW}^2}{2\Lambda^2}.
\end{aligned}\end{equation*}

\begin{figure}[!hbt]
  \centering
    \includegraphics[width=0.49\textwidth]{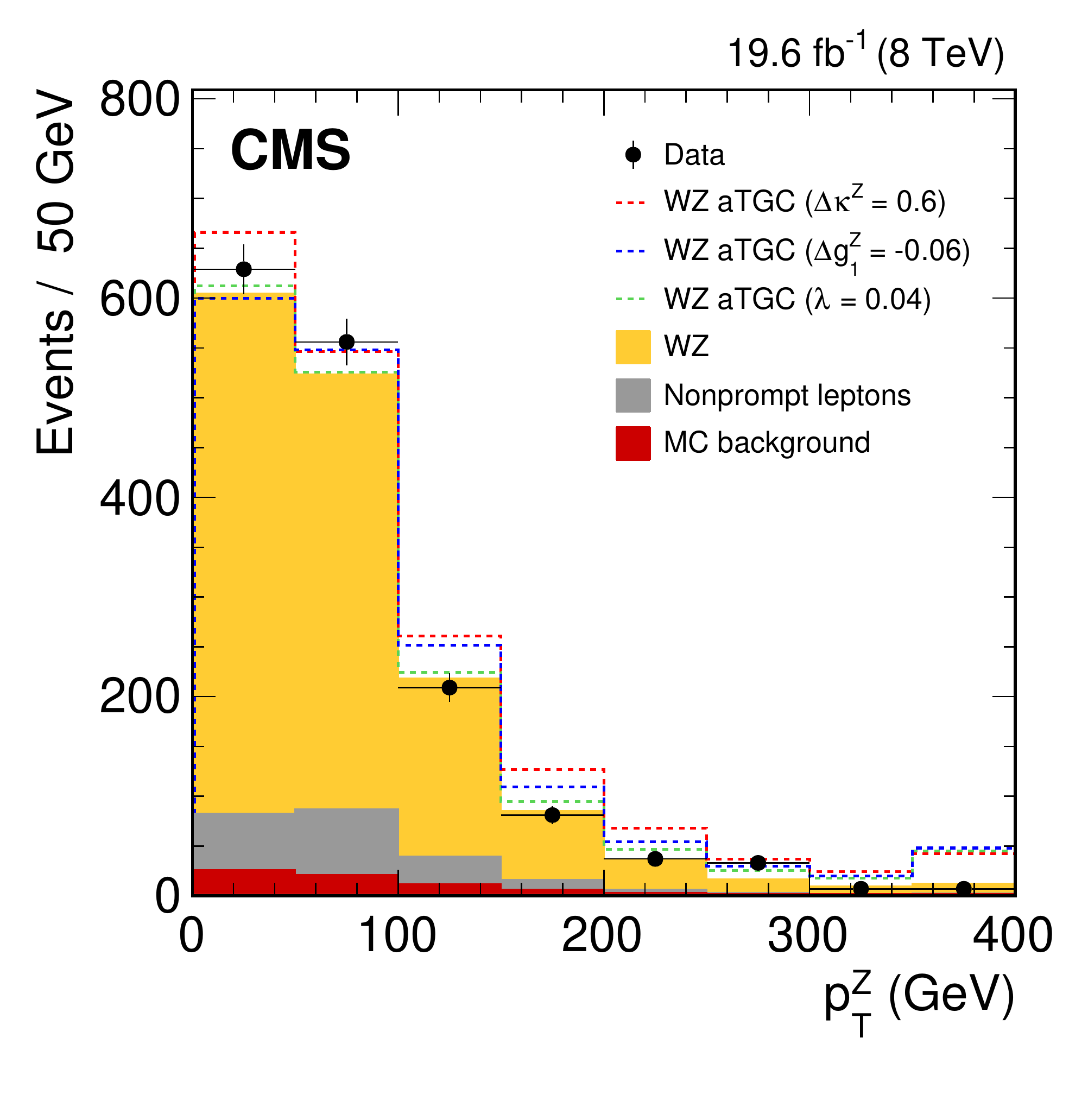}
    \includegraphics[width=0.49\textwidth]{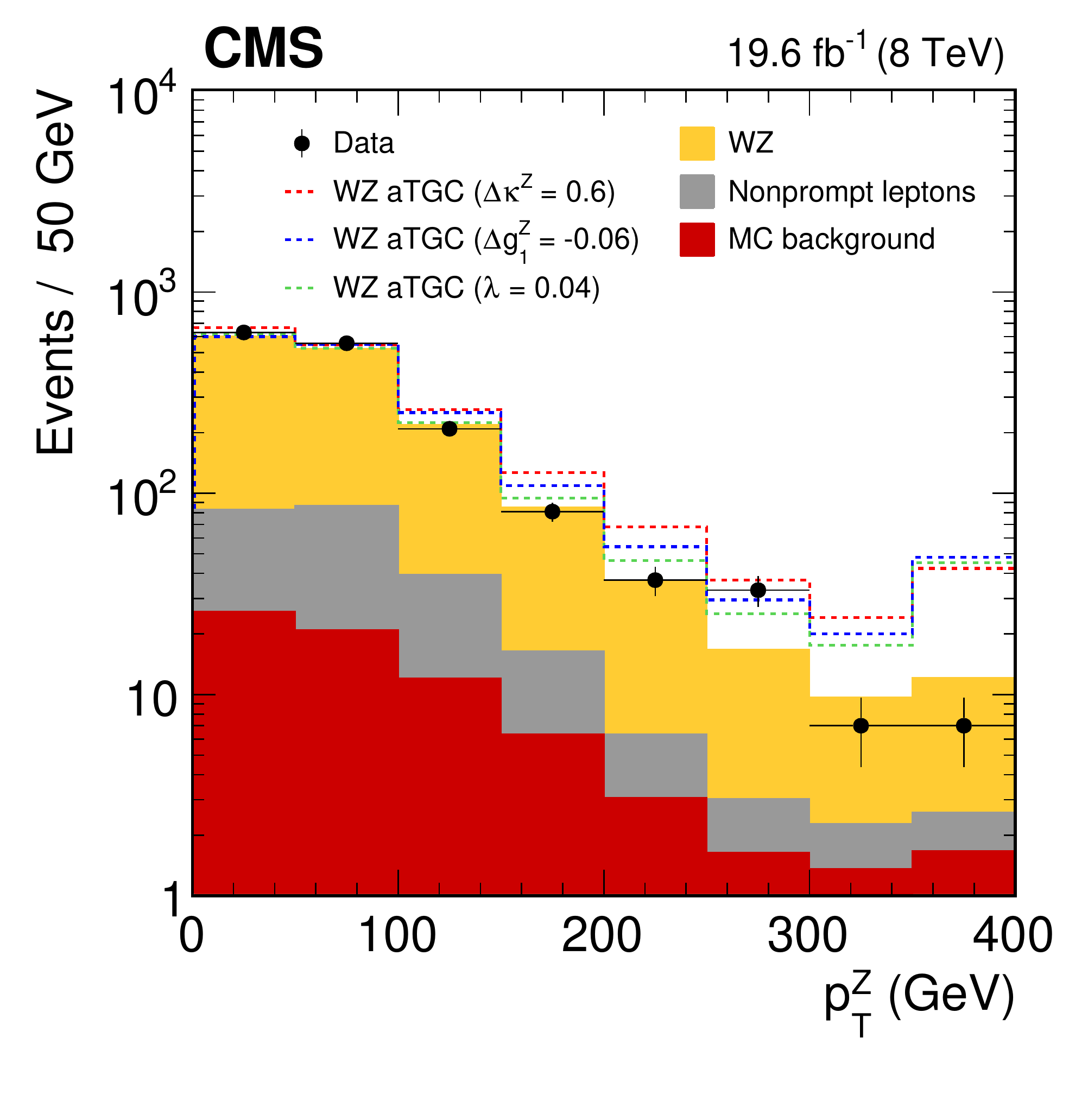}
    \caption{Transverse momentum distribution of the Z boson candidates, in linear scale
    (\cmsLeft) and log scale (\cmsRight) for all channels combined. The SM WZ contribution (light
    orange) is normalized to the predicted cross section from \MCFM. Dashed lines
    correspond to aTGC expectations with different parameter values. The last
    bin includes the integral of the tail. \label{fig:atgc_ptz}}
\end{figure}

\begin{figure}[hbt]
  \centering
    \includegraphics[width=\cmsFigWidth]{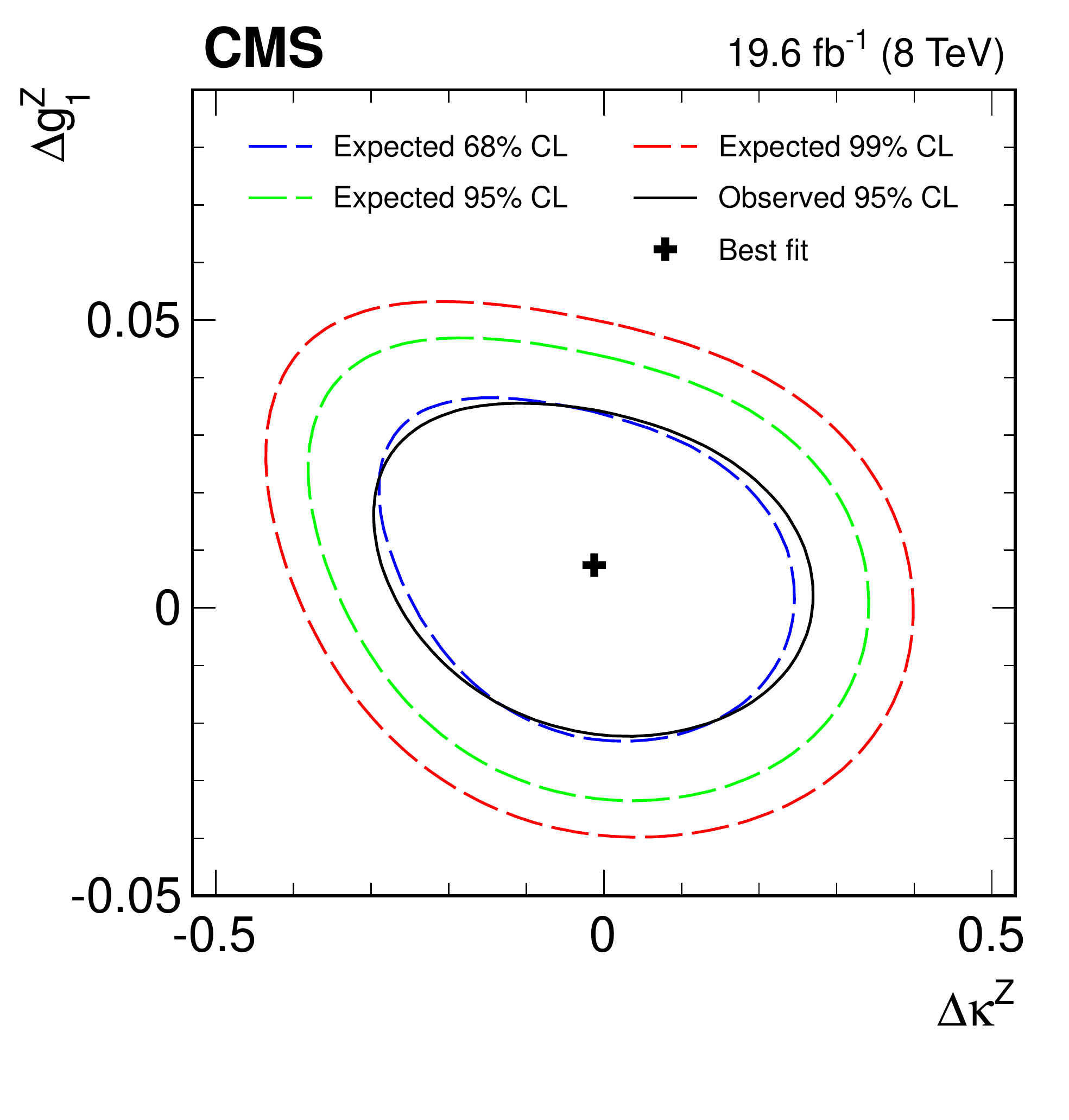}
   \caption{Two-dimensional observed 95\% CL limits and expected 68\%,
     95\% and 99\% CL limits on anomalous
    coupling parameters $\Delta\kappa^{\Z}$ and
    $\Delta g_{1}^{\Z}$. \label{fig:atgc_2dlimit_lam0}}
\end{figure}

\begin{figure}[hbt]
  \centering
   \includegraphics[width=\cmsFigWidth]{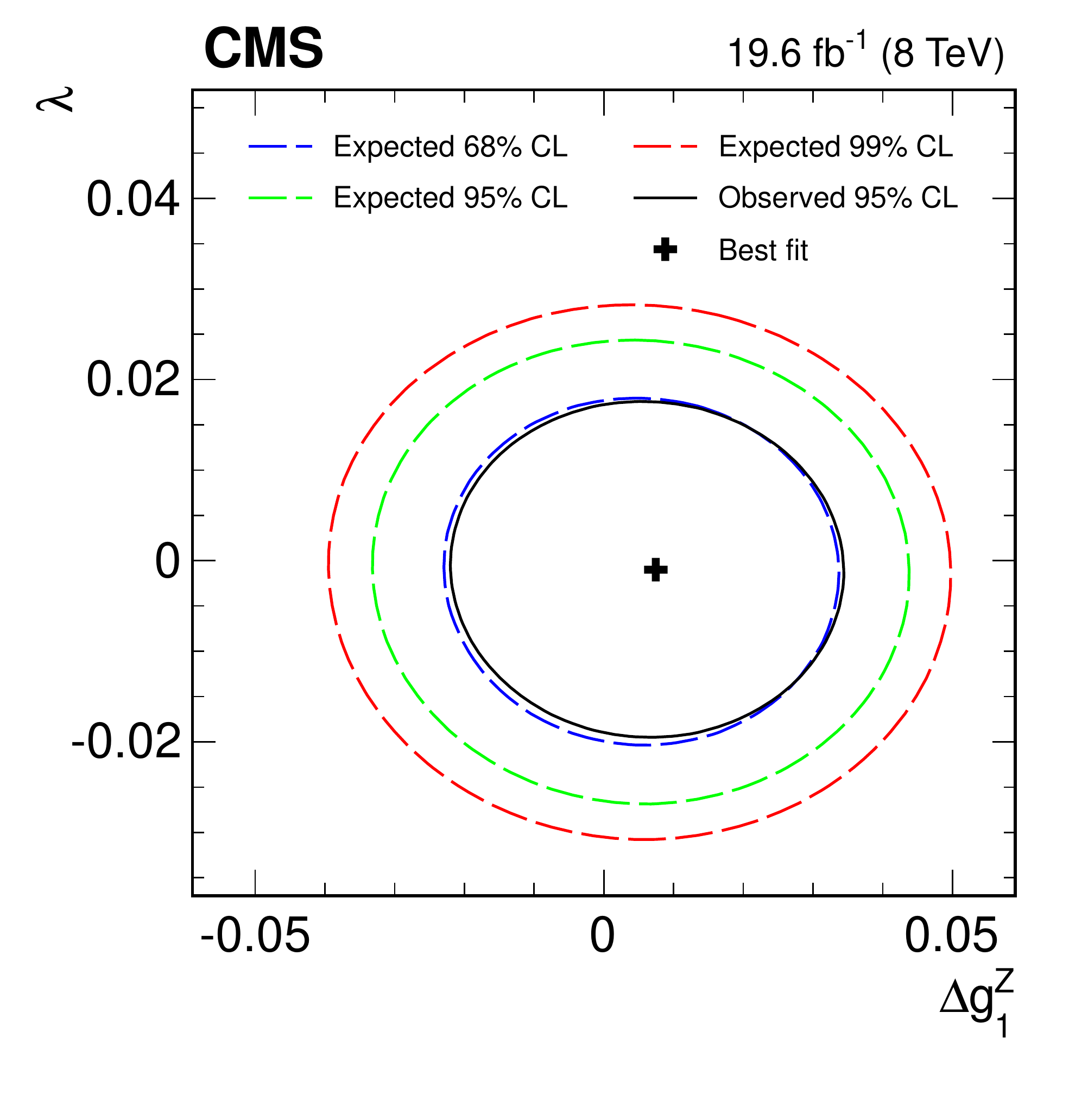}
   \caption{Two-dimensional observed 95\% CL limits and expected 68\%,
     95\% and 99\% CL limits on anomalous
    coupling parameters     $\Delta g_{1}^{\Z}$ and
$\lambda^{\Z}$.\label{fig:atgc_2dlimit_dk0}}
\end{figure}

\begin{figure}[hbt]
  \centering
    \includegraphics[width=\cmsFigWidth]{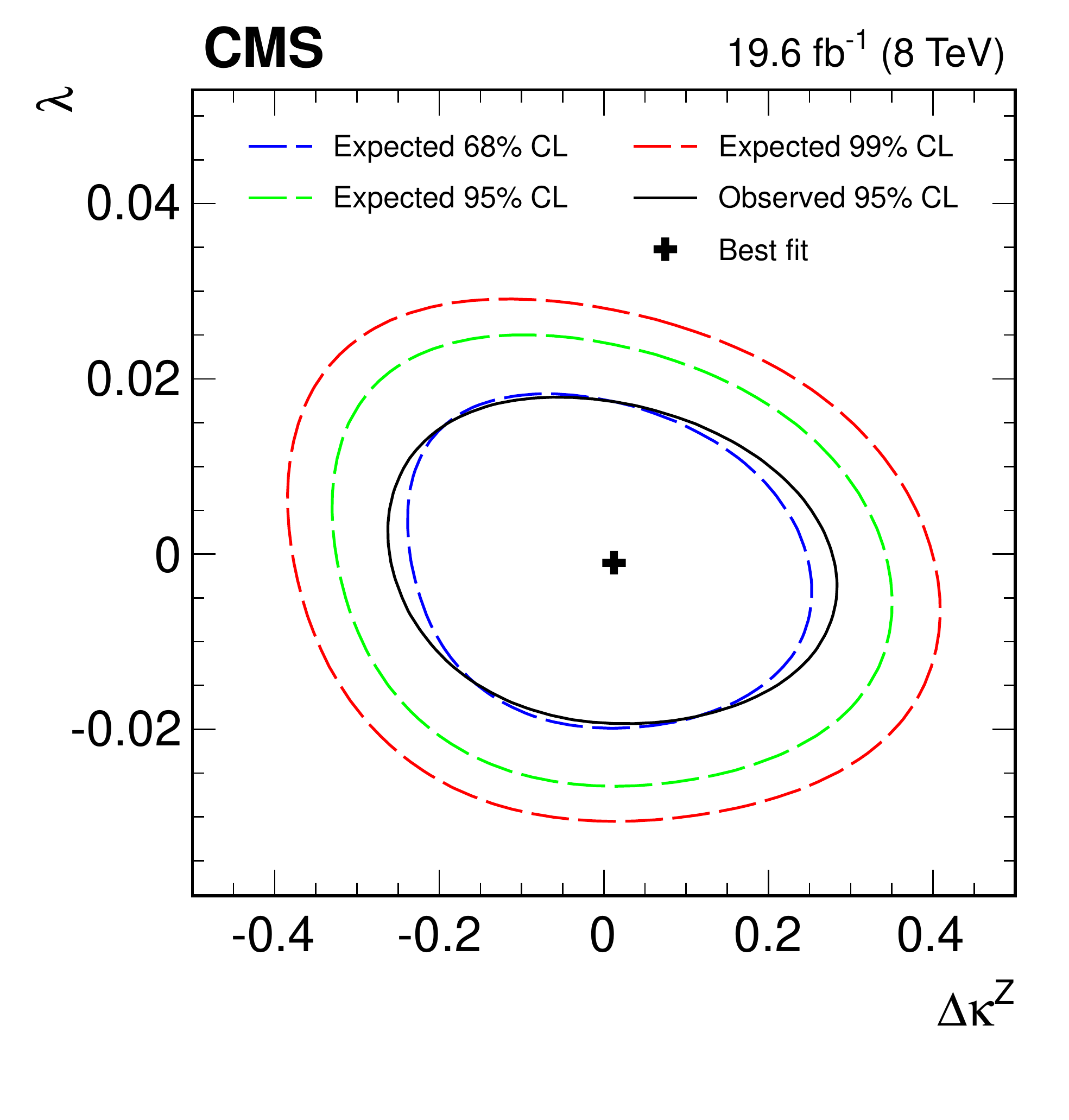}
   \caption{Two-dimensional observed 95\% CL limits and expected 68\%,
     95\% and 99\% CL limits on anomalous   coupling parameters
    $\Delta\kappa^{\Z}$ and $\lambda^{\Z}$.\label{fig:atgc_2dlimit_dg0}}
\end{figure}

The presence of anomalous triple gauge couplings would be manifested  as an
increased yield of events, with the largest increase at high Z boson
transverse momentum ($\pt^{\Z}$).
The expected $\pt^{\Z}$ spectrum for some aTGC values is
obtained by normalizing the \MADGRAPH  events to
the expected NLO SM cross section from \MCFM, and then reweighting
them to the expected cross section for that particular aTGC scenario,
as obtained with MCFM, based on the generated value of $\pt^{\Z}$.
Samples for three 2D anomalous parameter grids are generated,
$\lambda$ versus $\Delta\kappa^{\Z}$, $\lambda$ versus $\Delta g_{1}^{\Z}$,
and $\Delta\kappa^{\Z}$ versus $\Delta g_{1}^{\Z}$, where the
third parameter is set to its SM value. The expected yield of
the anomalous coupling signal in every $\pt^{\Z}$ bin is parametrized
by a second-order polynomial as a function of two aTGC parameters for every channel.
The observed $\pt^{\Z}$ spectrum is shown in
Fig.~\ref{fig:atgc_ptz} together with the expected spectra for a few
different aTGC scenarios.
A simultaneous fit to the values of aTGCs is performed~\cite{combine} in all four lepton
channels. A profile
 likelihood method, Wald gaussian approximation, and Wilks' theorem ~\cite{CCGV} are used to derive 1D and 2D
limits at a 95\% confidence level (CL) on each of the three aTGC
parameters and every combination of two aTGC parameters, respectively,
while all other parameters are set to their SM values.
 No significant deviation from the SM expectation is observed. Results can be found in Tables~\ref{tab:WgammaAtgcLimitsN0_1}
and~\ref{tab:WgammaAtgcLimitsN0_2}, and in
Figs.~\ref{fig:atgc_2dlimit_lam0}, \ref{fig:atgc_2dlimit_dk0}, and~\ref{fig:atgc_2dlimit_dg0}.

Limits on aTGC parameters were previously set by LEP~\cite{Schael:2013ita},
ATLAS~\cite{Aad:2014mda, Aad:2016ett} and CMS~\cite{CMS_WV7TeV:aTGC}. LHC analyses using 8\TeV data are setting most stringent limits. Results in this paper show sensitivity similar to the results given by the ATLAS Collaboration in the same channel~\cite{Aad:2016ett}.

Following the calculation in Ref.~\cite{Corbett:2014ora} we find the
lowest incoming parton energy for which observed limits on the
coefficients would lead to unitarity violation
(Table~\ref{tab:WgammaAtgcLimitsN0_v0_unitarity}).
Overall, for charged aTGCs, we are in the region where unitarity is not violated.

\begin{table*}[!htb]
\centering
\topcaption{One-dimensional limits on the aTGC parameters at a 95\% CL
for $\PW\Z\to\ell \nu \ell'\ell'$.\label{tab:WgammaAtgcLimitsN0_1}}
\begin{tabular}{l|cc}
\hline
                        &       Observed        & Expected \\ \hline
$\Delta\kappa^{\Z}$ &     [$-0.21,  0.25$]  &  [$-0.29,  0.30$] \\
$\Delta g_{1}^{\Z}$ &     [$-0.018, 0.035$] &  [$-0.028, 0.040$] \\
$\lambda^{\Z}$      &     [$-0.018, 0.016$] &  [$-0.024, 0.021$]\\
\hline
\end{tabular}
\end{table*}

\begin{table*}[!htb]
\centering
\topcaption{One-dimensional limits on the EFT parameters at a 95\% CL
for ${\PW\Z}\to \ell \nu \ell'\ell'$.\label{tab:WgammaAtgcLimitsN0_2}}
\begin{tabular}{l|cc}
\hline
                          & Observed [$\TeVns^{-2}$] & Expected [$\TeVns^{-2}$] \\ \hline
$c_{\PB}/\Lambda^{2}$   & [$-260, 210$] &  [$-310, 300$] \\
$c_{\PW}/\Lambda^{2}$   & [$-4.2, 8.0$] &  [$-6.8, 9.2$] \\
$c_{\PW\PW\PW}/\Lambda^{2}$ & [$-4.6, 4.2$] &  [$-6.1, 5.6$] \\
\hline
\end{tabular}
\end{table*}

\begin{table}[!htb]
\topcaption{Lowest incoming partons energy for which observed limits on the coefficients would lead to unitarity violation. \label{tab:WgammaAtgcLimitsN0_v0_unitarity}}
\centering
\begin{tabular}{l|cc} \hline
                   &       $\sqrt{s}$ [\TeVns{}]     \\
  \hline
  From observed limit on $c_{\PB}/\Lambda^{2}$ parameter  &   1.6      \\
  From observed limit on $c_{\PW}/\Lambda^{2}$ parameter  &   5.1      \\
  From observed limit on $c_{\PW\PW\PW}/\Lambda^{2}$ parameter &  4.3       \\
\hline
\end{tabular}
\end{table}

\section{Summary}

{\tolerance=1200
This paper reports measurements of the WZ inclusive
cross section in proton-proton collisions at
$\sqrt{s} = 7$ and 8\TeV
in the fully-leptonic WZ decay modes with electrons and muons in the final state.
The data samples correspond to integrated
luminosities of 4.9\fbinv for the 7\TeV measurement and 19.6\fbinv for the
8\TeV measurement. The measured production cross sections for
$71 < m_{\Z} < 111\GeV$ are
$\sigma({\Pp\Pp}\to{\PW\Z};~\sqrt{s} = 7\TeV) = 20.14 \pm 1.32\stat \pm  0.38\thy\pm 1.06\exper  \pm 0.44\lum$\unit{pb}
and
$\sigma({\Pp\Pp}\to{\PW\Z};~\sqrt{s} = 8\TeV) = 24.09 \pm 0.87\stat \pm  0.80\thy\pm 1.40\exper \pm 0.63\lum$\unit{pb}.
These results
are consistent with standard model predictions.

Using the data collected at $\sqrt{s} = 8\TeV$, results on differential
cross sections are also presented, and a search for
anomalous WWZ couplings has been performed. The
following one-dimensional limits at 95\% CL are obtained:
$-0.21 < \Delta\kappa^{\Z} < 0.25$, $-0.018 < \Delta g_{1}^{\Z} < 0.035$,
and $-0.018 < \lambda^{\Z} < 0.016$.
\par}

\clearpage
\begin{acknowledgments}
We congratulate our colleagues in the CERN accelerator departments for the excellent performance of the LHC and thank the technical and administrative staffs at CERN and at other CMS institutes for their contributions to the success of the CMS effort. In addition, we gratefully acknowledge the computing centres and personnel of the Worldwide LHC Computing Grid for delivering so effectively the computing infrastructure essential to our analyses. Finally, we acknowledge the enduring support for the construction and operation of the LHC and the CMS detector provided by the following funding agencies: BMWFW and FWF (Austria); FNRS and FWO (Belgium); CNPq, CAPES, FAPERJ, and FAPESP (Brazil); MES (Bulgaria); CERN; CAS, MoST, and NSFC (China); COLCIENCIAS (Colombia); MSES and CSF (Croatia); RPF (Cyprus); SENESCYT (Ecuador); MoER, ERC IUT and ERDF (Estonia); Academy of Finland, MEC, and HIP (Finland); CEA and CNRS/IN2P3 (France); BMBF, DFG, and HGF (Germany); GSRT (Greece); OTKA and NIH (Hungary); DAE and DST (India); IPM (Iran); SFI (Ireland); INFN (Italy); MSIP and NRF (Republic of Korea); LAS (Lithuania); MOE and UM (Malaysia); BUAP, CINVESTAV, CONACYT, LNS, SEP, and UASLP-FAI (Mexico); MBIE (New Zealand); PAEC (Pakistan); MSHE and NSC (Poland); FCT (Portugal); JINR (Dubna); MON, RosAtom, RAS and RFBR (Russia); MESTD (Serbia); SEIDI and CPAN (Spain); Swiss Funding Agencies (Switzerland); MST (Taipei); ThEPCenter, IPST, STAR and NSTDA (Thailand); TUBITAK and TAEK (Turkey); NASU and SFFR (Ukraine); STFC (United Kingdom); DOE and NSF (USA).

Individuals have received support from the Marie-Curie programme and the European Research Council and EPLANET (European Union); the Leventis Foundation; the A. P. Sloan Foundation; the Alexander von Humboldt Foundation; the Belgian Federal Science Policy Office; the Fonds pour la Formation \`a la Recherche dans l'Industrie et dans l'Agriculture (FRIA-Belgium); the Agentschap voor Innovatie door Wetenschap en Technologie (IWT-Belgium); the Ministry of Education, Youth and Sports (MEYS) of the Czech Republic; the Council of Science and Industrial Research, India; the HOMING PLUS programme of the Foundation for Polish Science, cofinanced from European Union, Regional Development Fund, the Mobility Plus programme of the Ministry of Science and Higher Education, the National Science Center (Poland), contracts Harmonia 2014/14/M/ST2/00428, Opus 2013/11/B/ST2/04202, 2014/13/B/ST2/02543 and 2014/15/B/ST2/03998, Sonata-bis 2012/07/E/ST2/01406; the Thalis and Aristeia programmes cofinanced by EU-ESF and the Greek NSRF; the National Priorities Research Program by Qatar National Research Fund; the Programa Clar\'in-COFUND del Principado de Asturias; the Rachadapisek Sompot Fund for Postdoctoral Fellowship, Chulalongkorn University and the Chulalongkorn Academic into Its 2nd Century Project Advancement Project (Thailand); and the Welch Foundation, contract C-1845.
\end{acknowledgments}

\bibliography{auto_generated}

\providecommand{\href}[2]{#2}\begingroup\raggedright\begin{thebibliography}{10}%
\makeatletter
\providecommand{\hrefCMSnoop }[0]{\@secondoftwo}%
\makeatother
\providecommand{\doi}{\texttt{doi:}\begingroup \urlstyle{tt}\Url}

\bibitem{Chatrchyan:2013iaa}
\hrefCMSnoop {}{{CMS Collaboration}, ``{Measurement of Higgs boson production
  and properties in the WW decay channel with leptonic final states}'',}
  \textit{ JHEP} \textbf{ 01} (2014) 096,
  \href{http://dx.doi.org/10.1007/JHEP01(2014)096}{\doi{10.1007/JHEP01(2014)096}},
\href{http://www.arXiv.org/abs/1312.1129}{\texttt{arXiv:1312.1129}}.

\bibitem{Khachatryan:2014xja}
\hrefCMSnoop {}{{CMS Collaboration}, ``Search for new resonances decaying via
  {WZ} to leptons in proton-proton collisions at {$\sqrt{s} = 8\TeV$}'',}
  \textit{ Phys. Lett. B} \textbf{ 740} (2015) 83,
  \href{http://dx.doi.org/10.1016/j.physletb.2014.11.026}{\doi{10.1016/j.physletb.2014.11.026}},
\href{http://www.arXiv.org/abs/1407.3476}{\texttt{arXiv:1407.3476}}.

\bibitem{Aad:2014pha}
\hrefCMSnoop {}{{ATLAS Collaboration}, ``Search for {$WZ$} resonances in the
  fully leptonic channel using $pp$ collisions at {$\sqrt{s}= 8\TeV$} with the
  {ATLAS} detector'',} \textit{ Phys. Lett. B} \textbf{ 737} (2014) 223,
  \href{http://dx.doi.org/10.1016/j.physletb.2014.08.039}{\doi{10.1016/j.physletb.2014.08.039}},
\href{http://www.arXiv.org/abs/1406.4456}{\texttt{arXiv:1406.4456}}.

\bibitem{Khachatryan:2014qwa}
\hrefCMSnoop {}{{CMS Collaboration}, ``Searches for electroweak production of
  charginos, neutralinos, and sleptons decaying to leptons and {W}, {Z}, and
  {H}iggs bosons in pp collisions at 8 {TeV}'',} \textit{ Eur. Phys. J. C}
  \textbf{ 74} (2014), no.~9, 3036,
  \href{http://dx.doi.org/10.1140/epjc/s10052-014-3036-7}{\doi{10.1140/epjc/s10052-014-3036-7}},
\href{http://www.arXiv.org/abs/1405.7570}{\texttt{arXiv:1405.7570}}.

\bibitem{Chatrchyan:2014aea}
\hrefCMSnoop {}{{CMS Collaboration}, ``Search for anomalous production of
  events with three or more leptons in $pp$ collisions at {$\sqrt(s) =
  8\TeV$}'',} \textit{ Phys. Rev. D} \textbf{ 90} (2014) 032006,
  \href{http://dx.doi.org/10.1103/PhysRevD.90.032006}{\doi{10.1103/PhysRevD.90.032006}},
\href{http://www.arXiv.org/abs/1404.5801}{\texttt{arXiv:1404.5801}}.

\bibitem{Aad:2016tuk}
\hrefCMSnoop {}{{ATLAS Collaboration}, ``Search for supersymmetry at
  {$\sqrt{s}=13\TeV$} in final states with jets and two same-sign leptons or
  three leptons with the {ATLAS} detector'',} \textit{ Eur. Phys. J. C}
  \textbf{ 76} (2016), no.~5, 259,
  \href{http://dx.doi.org/10.1140/epjc/s10052-016-4095-8}{\doi{10.1140/epjc/s10052-016-4095-8}},
\href{http://www.arXiv.org/abs/1602.09058}{\texttt{arXiv:1602.09058}}.

\bibitem{Aad:2011cwa}
\hrefCMSnoop {}{{ATLAS Collaboration}, ``Searches for supersymmetry with the
  {ATLAS} detector using final states with two leptons and missing transverse
  momentum in {$\sqrt{s}=7\TeV$} proton-proton collisions'',} \textit{ Phys.
  Lett. B} \textbf{ 709} (2012) 137,
  \href{http://dx.doi.org/10.1016/j.physletb.2012.01.076}{\doi{10.1016/j.physletb.2012.01.076}},
\href{http://www.arXiv.org/abs/1110.6189}{\texttt{arXiv:1110.6189}}.

\bibitem{Abazov:2012cj}
\hrefCMSnoop {}{{D0} Collaboration, ``{A measurement of the WZ and ZZ
  production cross sections using leptonic final states in 8.6 fb$^{-1}$ of
  $p\bar{p}$ collisions}'',} \textit{ Phys. Rev. D} \textbf{ 85} (2012) 112005,
  \href{http://dx.doi.org/10.1103/PhysRevD.85.112005}{\doi{10.1103/PhysRevD.85.112005}},
\href{http://www.arXiv.org/abs/1201.5652}{\texttt{arXiv:1201.5652}}.

\bibitem{Aaltonen:2012vu}
\hrefCMSnoop {}{{CDF} Collaboration, ``Measurement of the {WZ} cross section
  and triple gauge couplings in $p\bar{p}$ collisions at {$\sqrt{s} =
  1.96$~TeV}'',} \textit{ Phys. Rev. D} \textbf{ 86} (2012) 031104,
  \href{http://dx.doi.org/10.1103/PhysRevD.86.031104}{\doi{10.1103/PhysRevD.86.031104}},
\href{http://www.arXiv.org/abs/1202.6629}{\texttt{arXiv:1202.6629}}.

\bibitem{Aad:2012twa}
\hrefCMSnoop {}{{ATLAS Collaboration}, ``{Measurement of $WZ$ production in
  proton-proton collisions at $\sqrt{s}=7$ TeV with the ATLAS detector}'',}
  \textit{ Eur. Phys. J. C} \textbf{ 72} (2012) 2173,
  \href{http://dx.doi.org/10.1140/epjc/s10052-012-2173-0}{\doi{10.1140/epjc/s10052-012-2173-0}},
\href{http://www.arXiv.org/abs/1208.1390}{\texttt{arXiv:1208.1390}}.

\bibitem{Aad:2016ett}
\hrefCMSnoop {}{{ATLAS Collaboration}, ``{Measurements of $W^\pm Z$ production
  cross sections in $pp$ collisions at $\sqrt{s} = 8$ TeV with the ATLAS
  detector and limits on anomalous gauge boson self-couplings}'',} \textit{
  Phys. Rev. D} \textbf{ 93} (2016) 092004,
  \href{http://dx.doi.org/10.1103/PhysRevD.93.092004}{\doi{10.1103/PhysRevD.93.092004}},
\href{http://www.arXiv.org/abs/1603.02151}{\texttt{arXiv:1603.02151}}.

\bibitem{Aaboud:2016yus}
\hrefCMSnoop {}{{ATLAS Collaboration}, ``{Measurement of the
  $\rm{W}^{\pm}\rm{Z}$-boson production cross sections in pp collisions at
  $\sqrt{s}=13$ TeV with the ATLAS detector}'',} \textit{ Phys. Lett. B}
  \textbf{ 759} (2016) 601,
  \href{http://dx.doi.org/10.1016/j.physletb.2016.06.023}{\doi{10.1016/j.physletb.2016.06.023}},
\href{http://www.arXiv.org/abs/1606.04017}{\texttt{arXiv:1606.04017}}.

\bibitem{Khachatryan:2016tgp}
\hrefCMSnoop {}{{CMS Collaboration}, ``Measurement of the {WZ} production cross
  section in pp collisions at {$\sqrt{s} = 13\TeV$}'',} \textit{ Phys. Lett. B}
  \textbf{ 766} (2017) 268,
  \href{http://dx.doi.org/10.1016/j.physletb.2017.01.011}{\doi{10.1016/j.physletb.2017.01.011}},
\href{http://www.arXiv.org/abs/1607.06943}{\texttt{arXiv:1607.06943}}.

\bibitem{Aad:2014mda}
\hrefCMSnoop {}{{ATLAS Collaboration}, ``{Measurement of the WW+WZ cross
  section and limits on anomalous triple gauge couplings using final states
  with one lepton, missing transverse momentum, and two jets with the ATLAS
  detector at $\sqrt{\rm{s}} = 7$ TeV}'',} \textit{ JHEP} \textbf{ 01} (2015)
  049,
  \href{http://dx.doi.org/10.1007/JHEP01(2015)049}{\doi{10.1007/JHEP01(2015)049}},
\href{http://www.arXiv.org/abs/1410.7238}{\texttt{arXiv:1410.7238}}.

\bibitem{CMS_WV7TeV:aTGC}
\hrefCMSnoop {}{{CMS Collaboration}, ``Measurement of the sum of {WW} and {WZ}
  production with {W}+dijet events in pp collisions at {$\sqrt{s} = 7$\TeV}'',}
  \textit{ Eur. Phys. J. C} \textbf{ 73} (2013) 2283,
  \href{http://dx.doi.org/10.1140/epjc/s10052-013-2283-3}{\doi{10.1140/epjc/s10052-013-2283-3}}.

\bibitem{Agashe:2014kda}
\hrefCMSnoop {}{{Particle Data Group}, K.~A. Olive {et~al.}, ``{Review of
  Particle Physics}'',} \textit{ Chin. Phys. C} \textbf{ 38} (2014) 090001,
\href{http://dx.doi.org/10.1088/1674-1137/38/9/090001}{\doi{10.1088/1674-1137/38/9/090001}}.

\bibitem{Chatrchyan:2008zzk}
\hrefCMSnoop {}{{CMS Collaboration}, ``The {CMS} experiment at the {CERN}
  {LHC}'',} \textit{ JINST} \textbf{ 3} (2008) S08004,
\href{http://dx.doi.org/10.1088/1748-0221/3/08/S08004}{\doi{10.1088/1748-0221/3/08/S08004}}.

\bibitem{Madgraph5}
J.~Alwall\hrefCMSnoop {}{ {et~al.}, ``MadGraph 5: going beyond'',} \textit{
  JHEP} \textbf{ 06} (2011) 128,
  \href{http://dx.doi.org/10.1007/JHEP06(2011)128}{\doi{10.1007/JHEP06(2011)128}},
\href{http://www.arXiv.org/abs/1106.0522}{\texttt{arXiv:1106.0522}}.

\bibitem{Alioli:2008gx}
\hrefCMSnoop {}{S.~Alioli, P.~Nason, C.~Oleari, and E.~Re, ``{NLO vector-boson
  production matched with shower in POWHEG}'',} \textit{ JHEP} \textbf{ 07}
  (2008) 060,
  \href{http://dx.doi.org/10.1088/1126-6708/2008/07/060}{\doi{10.1088/1126-6708/2008/07/060}},
\href{http://www.arXiv.org/abs/0805.4802}{\texttt{arXiv:0805.4802}}.

\bibitem{Nason:2004rx}
\hrefCMSnoop {}{P.~Nason, ``{A new method for combining NLO QCD with shower
  Monte Carlo algorithms}'',} \textit{ JHEP} \textbf{ 11} (2004) 040,
  \href{http://dx.doi.org/10.1088/1126-6708/2004/11/040}{\doi{10.1088/1126-6708/2004/11/040}},
\href{http://www.arXiv.org/abs/hep-ph/0409146}{\texttt{arXiv:hep-ph/0409146}}.

\bibitem{Frixione:2007vw}
\hrefCMSnoop {}{S.~Frixione, P.~Nason, and C.~Oleari, ``{Matching NLO QCD
  computations with parton shower simulations: the POWHEG method}'',} \textit{
  JHEP} \textbf{ 11} (2007) 070,
  \href{http://dx.doi.org/10.1088/1126-6708/2007/11/070}{\doi{10.1088/1126-6708/2007/11/070}},
\href{http://www.arXiv.org/abs/0709.2092}{\texttt{arXiv:0709.2092}}.

\bibitem{Binoth:2008pr}
\hrefCMSnoop {}{T.~Binoth, N.~Kauer, and P.~Mertsch, ``{Gluon-induced QCD
  corrections to pp $\to$ ZZ $\to \ell\bar{\ell}\ell'\bar{\ell'}$}'',} in
  \textit{ {Proceedings, 16th International Workshop on Deep Inelastic
  Scattering and Related Subjects (DIS 2008)}}, p.~142.
\newblock 2008.
\newblock \href{http://www.arXiv.org/abs/0807.0024}{\texttt{arXiv:0807.0024}}.
\newblock
\href{http://dx.doi.org/10.3360/dis.2008.142}{\doi{10.3360/dis.2008.142}}.

\bibitem{MCFM}
\hrefCMSnoop {}{J.~M. Campbell and R.~K. Ellis, ``{MCFM for the Tevatron and
  the LHC}'',} \textit{ Nucl. Phys. Proc. Suppl.} \textbf{ 205-206} (2010) 10,
  \href{http://dx.doi.org/10.1016/j.nuclphysbps.2010.08.011}{\doi{10.1016/j.nuclphysbps.2010.08.011}},
\href{http://www.arXiv.org/abs/1007.3492}{\texttt{arXiv:1007.3492}}.

\bibitem{Sjostrand:2006za}
\hrefCMSnoop {}{T.~Sj{\"o}strand, S.~Mrenna, and P.~Z. Skands, ``{PYTHIA} 6.4
  physics and manual'',} \textit{ JHEP} \textbf{ 05} (2006) 026,
  \href{http://dx.doi.org/10.1088/1126-6708/2006/05/026}{\doi{10.1088/1126-6708/2006/05/026}},
\href{http://www.arXiv.org/abs/hep-ph/0603175}{\texttt{arXiv:hep-ph/0603175}}.

\bibitem{Field:2010bc}
\href {http://inspirehep.net/record/873443/files/arXiv:1010.3558.pdf}{R.~Field,
  ``{Early LHC Underlying Event Data -- Findings and Surprises}'',} in \textit{
  {Hadron collider physics. Proceedings, 22nd Conference, HCP 2010, Toronto,
  Canada, August 23-27, 2010}}.
\newblock 2010.
\newblock
\href{http://www.arXiv.org/abs/1010.3558}{\texttt{arXiv:1010.3558}}.
\newblock

\bibitem{CTEQ66}
H.-L. Lai\hrefCMSnoop {}{ {et~al.}, ``Uncertainty induced by {QCD} coupling in
  the {CTEQ} global analysis of parton distributions'',} \textit{ Phys. Rev. D}
  \textbf{ 82} (2010) 054021,
  \href{http://dx.doi.org/10.1103/PhysRevD.82.054021}{\doi{10.1103/PhysRevD.82.054021}},
  \href{http://www.arXiv.org/abs/1004.4624}{\texttt{arXiv:1004.4624}}.

\bibitem{ct10}
H.-L. Lai\hrefCMSnoop {}{ {et~al.}, ``New parton distributions for collider
  physics'',} \textit{ Phys. Rev. D} \textbf{ 82} (2010) 074024,
  \href{http://dx.doi.org/10.1103/PhysRevD.82.074024}{\doi{10.1103/PhysRevD.82.074024}},
  \href{http://www.arXiv.org/abs/1007.2241}{\texttt{arXiv:1007.2241}}.

\bibitem{GEANT}
\hrefCMSnoop {}{{GEANT4} Collaboration, ``GEANT4 --- a simulation toolkit'',}
  \textit{ Nucl. Instrum. Meth. A} \textbf{ 506} (2003) 250,
\href{http://dx.doi.org/10.1016/S0168-9002(03)01368-8}{\doi{10.1016/S0168-9002(03)01368-8}}.

\bibitem{CMS-PAS-PFT-09-001}
\href {http://cdsweb.cern.ch/record/1194487}{{CMS Collaboration},
  ``Particle--Flow Event Reconstruction in {CMS} and Performance for Jets,
  Taus, and $\vec{E}_{\rm T}^{\rm miss}$'',} CMS Physics Analysis Summary
  CMS-PAS-PFT-09-001, 2009.

\bibitem{CMS-PAS-PFT-10-001}
\href {http://cdsweb.cern.ch/record/1247373}{{CMS Collaboration},
  ``Commissioning of the Particle-flow Event Reconstruction with the first
  {LHC} collisions recorded in the {CMS} detector'',} CMS Physics Analysis
  Summary CMS-PAS-PFT-10-001, 2010.

\bibitem{Khachatryan:2015hwa}
\hrefCMSnoop {}{{CMS Collaboration}, ``Performance of electron reconstruction
  and selection with the {CMS} detector in proton-proton collisions at
  {$\sqrt{s} = 8$\TeV}'',} \textit{ JINST} \textbf{ 10} (2015) P06005,
  \href{http://dx.doi.org/10.1088/1748-0221/10/06/P06005}{\doi{10.1088/1748-0221/10/06/P06005}}.

\bibitem{CMS-PAS-MUO-10-002}
\hrefCMSnoop {}{{CMS Collaboration}, ``{Performance of CMS muon reconstruction
  in pp collision events at $\sqrt{s}=7\TeV$}'',} \textit{ JINST} \textbf{ 7}
  (2012) P10002,
  \href{http://dx.doi.org/10.1088/1748-0221/7/10/P10002}{\doi{10.1088/1748-0221/7/10/P10002}},
\href{http://www.arXiv.org/abs/1206.4071}{\texttt{arXiv:1206.4071}}.

\bibitem{Cacciari:subtraction}
\hrefCMSnoop {}{M.~Cacciari and G.~P. Salam, ``{Pileup subtraction using jet
  areas}'',} \textit{ Phys. Lett. B} \textbf{ 659} (2008) 119,
  \href{http://dx.doi.org/10.1016/j.physletb.2007.09.077}{\doi{10.1016/j.physletb.2007.09.077}},
  \href{http://www.arXiv.org/abs/0707.1378}{\texttt{arXiv:0707.1378}}.

\bibitem{wzxs}
\hrefCMSnoop {}{{CMS Collaboration}, ``Measurements of inclusive {W} and {Z}
  cross sections in pp collisions at $\sqrt{s}$ = 7 {TeV}'',} \textit{ JHEP}
  \textbf{ 01} (2011) 080,
  \href{http://dx.doi.org/10.1007/JHEP01(2011)080}{\doi{10.1007/JHEP01(2011)080}}.

\bibitem{Cacciari:2008gp}
\hrefCMSnoop {}{M.~Cacciari, G.~P. Salam, and G.~Soyez, ``The anti-$k_t$ jet
  clustering algorithm'',} \textit{ JHEP} \textbf{ 04} (2008) 063,
  \href{http://dx.doi.org/10.1088/1126-6708/2008/04/063}{\doi{10.1088/1126-6708/2008/04/063}},
  \href{http://www.arXiv.org/abs/0802.1189}{\texttt{arXiv:0802.1189}}.

\bibitem{Cacciari:2011ma}
\hrefCMSnoop {}{M.~Cacciari, G.~P. Salam, and G.~Soyez, ``{FastJet} user
  manual'',} \textit{ Eur. Phys. J. C} \textbf{ 72} (2012) 1896,
  \href{http://dx.doi.org/10.1140/epjc/s10052-012-1896-2}{\doi{10.1140/epjc/s10052-012-1896-2}},
\href{http://www.arXiv.org/abs/1111.6097}{\texttt{arXiv:1111.6097}}.

\bibitem{Chatrchyan:2011ds}
\hrefCMSnoop {}{{CMS Collaboration}, ``{Determination of jet energy calibration
  and transverse momentum resolution in CMS}'',} \textit{ JINST} \textbf{ 6}
  (2011) P11002,
  \href{http://dx.doi.org/10.1088/1748-0221/6/11/P11002}{\doi{10.1088/1748-0221/6/11/P11002}},
\href{http://www.arXiv.org/abs/1107.4277}{\texttt{arXiv:1107.4277}}.

\bibitem{CMSWWxsTGC35pb}
\hrefCMSnoop {}{{CMS Collaboration}, ``{Measurement of WW production and search
  for the Higgs boson in pp collisions at $\sqrt{s} = 7~\TeV$}'',} \textit{
  Phys. Lett. B} \textbf{ 699} (2011) 25,
  \href{http://dx.doi.org/10.1016/j.physletb.2011.03.056}{\doi{10.1016/j.physletb.2011.03.056}},
\href{http://www.arXiv.org/abs/1102.5429}{\texttt{arXiv:1102.5429}}.

\bibitem{pdfunc}
\hrefCMSnoop {}{J.~M. Campbell, W.~Huston, J, and W.~J. Stirling, ``Hard
  interactions of quarks and gluons: a primer for LHC physics'',} \textit{ Rep.
  Prog. Phys.} \textbf{ 70} (2007) 89,
  \href{http://dx.doi.org/10.1088/0034-4885/70/1/R02}{\doi{10.1088/0034-4885/70/1/R02}},
  \href{http://www.arXiv.org/abs/hep-ph/0611148}{\texttt{arXiv:hep-ph/0611148}}.

\bibitem{Chatrchyan:2012nj}
\hrefCMSnoop {}{{CMS Collaboration}, ``Measurement of the inelastic
  proton-proton cross section at {$\sqrt{s} = 7\TeV$}'',} \textit{ Phys. Lett.
  B} \textbf{ 722} (2013) 5,
  \href{http://dx.doi.org/10.1016/j.physletb.2013.03.024}{\doi{10.1016/j.physletb.2013.03.024}}.

\bibitem{Chatrchyan:2012sga}
\hrefCMSnoop {}{{CMS Collaboration}, ``Measurement of the ZZ production cross
  section and search for anomalous couplings in $2\ell2\ell'$ final states in
  pp collisions at {$\sqrt{s} = 7\TeV$}'',} \textit{ JHEP} \textbf{ 01} (2013)
  063,
  \href{http://dx.doi.org/10.1007/JHEP01(2013)063}{\doi{10.1007/JHEP01(2013)063}}.

\bibitem{Chatrchyan:2013oev}
\hrefCMSnoop {}{{CMS Collaboration}, ``Measurement of the
  {$\mathrm{W}^+\mathrm{W}^-$} and {ZZ} production cross sections in pp
  collisions at {$\sqrt{s} = 8\TeV$}'',} \textit{ Phys. Lett. B} \textbf{ 721}
  (2013) 190,
  \href{http://dx.doi.org/10.1016/j.physletb.2013.03.027}{\doi{10.1016/j.physletb.2013.03.027}}.

\bibitem{Chatrchyan:2013fya}
\hrefCMSnoop {}{{CMS Collaboration}, ``{Measurement of the $\PW\gamma$ and
  $\cPZ\gamma$ inclusive cross sections in $\Pp\Pp$ collisions at $\sqrt{s} =
  7\TeV$ and limits on anomalous triple gauge boson couplings}'',} \textit{
  Phys. Rev. D} \textbf{ 89} (2014) 092005,
  \href{http://dx.doi.org/10.1103/PhysRevD.89.092005}{\doi{10.1103/PhysRevD.89.092005}}.

\bibitem{Khachatryan:2015kea}
\hrefCMSnoop {}{{CMS Collaboration}, ``{Measurement of the Z$\gamma$ production
  cross section in pp collisions at 8 TeV and search for anomalous triple gauge
  boson couplings}'',} \textit{ JHEP} \textbf{ 04} (2015) 164,
  \href{http://dx.doi.org/10.1007/JHEP04(2015)164}{\doi{10.1007/JHEP04(2015)164}},
\href{http://www.arXiv.org/abs/1502.05664}{\texttt{arXiv:1502.05664}}.

\bibitem{Khachatryan:2014ewa}
\hrefCMSnoop {}{{CMS Collaboration}, ``Measurement of top quark-antiquark pair
  production in association with a {W} or {Z} boson in pp collisions at
  {$\sqrt{s} = 8\TeV$}'',} \textit{ Eur. Phys. J. C} \textbf{ 74} (2014) 3060,
  \href{http://dx.doi.org/10.1140/epjc/s10052-014-3060-7}{\doi{10.1140/epjc/s10052-014-3060-7}}.

\bibitem{CMS-PAS-SMP-12-008}
\href {https://cds.cern.ch/record/1434360}{{CMS Collaboration}, ``{Absolute
  Calibration of the Luminosity Measurement at CMS: Winter 2012 Update}'',} CMS
  Physics Analysis Summary CMS-PAS-SMP-12-008, 2012.

\bibitem{CMS-PAS-LUM-13-001}
\href {https://cds.cern.ch/record/1598864}{{CMS Collaboration}, ``{CMS
  Luminosity Based on Pixel Cluster Counting - Summer 2013 Update}'',} CMS
  Physics Analysis Summary CMS-PAS-LUM-13-001, 2013.

\bibitem{Lyons:1988rp}
\hrefCMSnoop {}{L.~Lyons, D.~Gibaut, and P.~Clifford, ``{How to combine
  correlated estimates of a single physical quantity}'',} \textit{ Nucl.
  Instrum. Meth. A} \textbf{ 270} (1988) 110,
  \href{http://dx.doi.org/10.1016/0168-9002(88)90018-6}{\doi{10.1016/0168-9002(88)90018-6}}.

\bibitem{Grazzini:2016swo}
\hrefCMSnoop {}{M.~Grazzini, S.~Kallweit, D.~Rathlev, and M.~Wiesemann,
  ``{$W^{\pm}Z$ production at hadron colliders in NNLO QCD}'',} \textit{ Phys.
  Lett. B} \textbf{ 761} (2016) 179,
  \href{http://dx.doi.org/10.1016/j.physletb.2016.08.017}{\doi{10.1016/j.physletb.2016.08.017}},
\href{http://www.arXiv.org/abs/1604.08576}{\texttt{arXiv:1604.08576}}.

\bibitem{D'Agostini:1994zf}
\hrefCMSnoop {}{G.~D'Agostini, ``{A multidimensional unfolding method based on
  Bayes' theorem}'',} \textit{ Nucl. Instrum. Meth. A} \textbf{ 362} (1995)
  487,
\href{http://dx.doi.org/10.1016/0168-9002(95)00274-X}{\doi{10.1016/0168-9002(95)00274-X}}.

\bibitem{RooUnfold}
\hrefCMSnoop {}{T.~Adye, ``Unfolding algorithms and tests using {RooUnfold}'',}
  in \textit{ {PHYSTAT} 2011 Workshop on Statistical Issues Related to
  Discovery Claims in Search Experiments and Unfolding}, H.~Prosper and
  L.~Lyons, eds., p.~313.
\newblock Geneva, Switzerland, 2011.
\newblock \href{http://www.arXiv.org/abs/1105.1160}{\texttt{arXiv:1105.1160}}.
\newblock
\href{http://dx.doi.org/10.5170/CERN-2011-006.313}{\doi{10.5170/CERN-2011-006.313}}.

\bibitem{Valassi:2003mu}
\hrefCMSnoop {}{A.~Valassi, ``Combining correlated measurements of several
  different physical quantities'',} \textit{ Nucl. Instrum. Meth. A} \textbf{
  500} (2003) 391,
\href{http://dx.doi.org/10.1016/S0168-9002(03)00329-2}{\doi{10.1016/S0168-9002(03)00329-2}}.

\bibitem{hagiwara1}
\hrefCMSnoop {}{K.~Hagiwara, R.~D. Peccei, and D.~Zeppenfeld, ``Probing the
  weak boson sector in {e$^+$e$^-\to {\rm{W}}^+ {\rm{W}}^-$}'',} \textit{ Nucl.
  Phys. B} \textbf{ 282} (1987) 253,
  \href{http://dx.doi.org/10.1016/0550-3213(87)90685-7}{\doi{10.1016/0550-3213(87)90685-7}}.

\bibitem{hagiwara2}
\hrefCMSnoop {}{K.~Hagiwara, J.~Woodside, and D.~Zeppenfeld, ``{Measuring the
  WWZ coupling at the Fermilab Tevatron}'',} \textit{ Phys. Rev. D} \textbf{
  41} (1990) 2113,
  \href{http://dx.doi.org/10.1103/PhysRevD.41.2113}{\doi{10.1103/PhysRevD.41.2113}}.

\bibitem{GrosseKnetter:1993rp}
\hrefCMSnoop {}{C.~Grosse-Knetter, I.~Kuss, and D.~Schildknecht, ``Nonstandard
  gauge boson selfinteractions within a gauge invariant model'',} \textit{ Z.
  Phys. C} \textbf{ 60} (1993) 375,
  \href{http://dx.doi.org/10.1007/BF01474637}{\doi{10.1007/BF01474637}},
\href{http://www.arXiv.org/abs/hep-ph/9304281}{\texttt{arXiv:hep-ph/9304281}}.

\bibitem{Bilenky:1993ms}
\hrefCMSnoop {}{M.~S. Bilenky, J.~L. Kneur, F.~M. Renard, and D.~Schildknecht,
  ``Trilinear couplings among the electroweak vector bosons and their
  determination at {LEP-200}'',} \textit{ Nucl. Phys. B} \textbf{ 409} (1993)
  22,
\href{http://dx.doi.org/10.1016/0550-3213(93)90445-U}{\doi{10.1016/0550-3213(93)90445-U}}.

\bibitem{Bilenky:1993uy}
\hrefCMSnoop {}{M.~S. Bilenky, J.-L. Kneur, F.~M. Renard, and D.~Schildknecht,
  ``The potential of a new linear collider for the measurement of the trilinear
  couplings among the electroweak vector bosons'',} \textit{ Nucl. Phys. B}
  \textbf{ 419} (1994) 240,
  \href{http://dx.doi.org/10.1016/0550-3213(94)90041-8}{\doi{10.1016/0550-3213(94)90041-8}},
\href{http://www.arXiv.org/abs/hep-ph/9312202}{\texttt{arXiv:hep-ph/9312202}}.

\bibitem{combine}
\href {http://cds.cern.ch/record/1379837}{{ATLAS and CMS Collaborations and LHC
  Higgs Combination Group}, ``{Procedure for the LHC Higgs boson search
  combination in Summer 2011}'',} CMS NOTE {CMS-NOTE-2011-005},
  {ATL-PHYS-PUB-2011-11}, CERN, Geneva, 2011.

\bibitem{CCGV}
\hrefCMSnoop {}{G.~Cowan, K.~Cranmer, E.~Gross, and O.~Vitells, ``Asymptotic
  formulae for likelihood-based tests of new physics'',} \textit{ Eur. Phys. J.
  C} \textbf{ 71} (2011) 1554,
  \href{http://dx.doi.org/10.1140/epjc/s10052-011-1554-0}{\doi{10.1140/epjc/s10052-011-1554-0}},
  \href{http://www.arXiv.org/abs/1007.1727}{\texttt{arXiv:1007.1727}}.

\bibitem{Schael:2013ita}
\hrefCMSnoop {}{{The ALEPH Collaboration, The DELPHI Collaboration, The L3
  Collaboration, The OPAL Collaboration, The LEP Electroweak Working Group},
  ``Electroweak measurements in electron-positron collisions at {W}-boson-pair
  energies at {LEP}'',} \textit{ Phys. Rept.} \textbf{ 532} (2013) 119,
  \href{http://dx.doi.org/10.1016/j.physrep.2013.07.004}{\doi{10.1016/j.physrep.2013.07.004}},
\href{http://www.arXiv.org/abs/1302.3415}{\texttt{arXiv:1302.3415}}.

\bibitem{Corbett:2014ora}
\hrefCMSnoop {}{T.~Corbett, O.~J.~P. Eboli, and M.~C. Gonzalez-Garcia,
  ``Unitarity constraints on dimension-six operators'',} \textit{ Phys. Rev. D}
  \textbf{ 91} (2015) 035014,
  \href{http://dx.doi.org/10.1103/PhysRevD.91.035014}{\doi{10.1103/PhysRevD.91.035014}},
\href{http://www.arXiv.org/abs/1411.5026}{\texttt{arXiv:1411.5026}}.

\end{thebibliography}\endgroup
\cleardoublepage \appendix\section{The CMS Collaboration \label{app:collab}}\begin{sloppypar}\hyphenpenalty=5000\widowpenalty=500\clubpenalty=5000\textbf{Yerevan Physics Institute,  Yerevan,  Armenia}\\*[0pt]
V.~Khachatryan, A.M.~Sirunyan, A.~Tumasyan
\vskip\cmsinstskip
\textbf{Institut f\"{u}r Hochenergiephysik der OeAW,  Wien,  Austria}\\*[0pt]
W.~Adam, E.~Asilar, T.~Bergauer, J.~Brandstetter, E.~Brondolin, M.~Dragicevic, J.~Er\"{o}, M.~Flechl, M.~Friedl, R.~Fr\"{u}hwirth\cmsAuthorMark{1}, V.M.~Ghete, C.~Hartl, N.~H\"{o}rmann, J.~Hrubec, M.~Jeitler\cmsAuthorMark{1}, A.~K\"{o}nig, I.~Kr\"{a}tschmer, D.~Liko, T.~Matsushita, I.~Mikulec, D.~Rabady, N.~Rad, B.~Rahbaran, H.~Rohringer, J.~Schieck\cmsAuthorMark{1}, J.~Strauss, W.~Treberer-Treberspurg, W.~Waltenberger, C.-E.~Wulz\cmsAuthorMark{1}
% ...
\vskip\cmsinstskip
% ...
1:~~Also at Vienna University of Technology, Vienna, Austria\\
% ...

\end{sloppypar}
\end{document}